\documentclass[prd,superscriptaddress,preprintnumbers,twocolumn,showpacs]{revtex4}
\usepackage{graphicx}
\usepackage{amsmath}
\usepackage{multirow}
\usepackage{bm}
\usepackage{color}
\usepackage{ulem}

\newcommand{\tr}{\,\mbox{tr}}
\newcommand{\sign}{\,\mbox{sign}}

\begin{document}
\normalem

\title{Normal ground state of dense relativistic matter in a magnetic field}
\date{January 25, 2011}

\preprint{UWO-TH-11/1}

\author{E. V. Gorbar}
\email{gorbar@bitp.kiev.ua}
\affiliation{Bogolyubov Institute for Theoretical Physics, 03680, Kiev, Ukraine}

\author{V. A. Miransky}
\email{vmiransk@uwo.ca}
\affiliation{Department of Applied Mathematics, University of Western Ontario, London, Ontario N6A 5B7, Canada}

\author{I. A. Shovkovy}
\email{igor.shovkovy@asu.edu}
\affiliation{Department of Applied Sciences and Mathematics, Arizona State University, Mesa, Arizona 85212, USA}

\begin{abstract}
The properties of the ground state of relativistic matter in a magnetic field are examined within the framework
 of a Nambu-Jona-Lasinio model. The main emphasis of this study is the normal ground state,
which is realized at sufficiently high temperatures and/or sufficiently large chemical potentials. In contrast to the 
vacuum state, which is characterized by the magnetic catalysis of chiral symmetry breaking, the
normal state is accompanied by the dynamical generation of the chiral shift parameter $\Delta$. In the chiral limit, 
the value of $\Delta$ determines a relative shift of the longitudinal momenta (along the direction of
the magnetic field) in the dispersion relations of opposite chirality fermions. We argue that the chirality remains a 
good approximate quantum number even for massive fermions in the vicinity of the Fermi surface and, therefore, the chiral 
shift is expected to play an important role in many types of cold dense relativistic matter, relevant for applications in 
compact stars. The qualitative implications of the revealed structure of the normal ground state on the physics of 
protoneutron stars are discussed. A noticeable feature of the $\Delta$ parameter 
is that it is insensitive to temperature when $T \ll \mu_0$, where $\mu_0$ is the chemical potential, 
and {\it increases} with temperature for $T > \mu_0$. The latter implies that the chiral shift parameter 
is also generated in the regime relevant for heavy ion collisions.
\end{abstract}


\maketitle

\section{Introduction}

Dense relativistic matter in strong magnetic fields naturally exists in compact stars. For example,
such type of matter is formed by the electron component of the nuclear matter in the interior of neutron
stars. At lower density, a relativistic electron plasma exists and plays an essential role in white dwarfs.
In both cases, magnetic fields could be rather strong: they reach up to $10^{9}~\mbox{G}$ in white dwarfs
and up to $10^{15}~\mbox{G}$ in neutron stars \cite{astroreview,astroreview1}. If quark stars 
exist in nature, the corresponding dense quark matter in the core will be a strongly coupled version of 
relativistic matter. Relativistic matter in a strong magnetic field is also created in heavy ion collisions 
\cite{Skokov:2009qp} that can lead to the chiral magnetic effect \cite{Kharzeev:2007tn,Buividovich:2009wi}.

Many physical properties of the stellar matter under extreme conditions realized inside compact
stars are understood theoretically and could be tested to some extent through observational
data. {However, as was pointed out in Refs.~\cite{FI1,Metlitski:2005pr,SS,Gorbar:2009bm,Rebhan,
Preis,Basar:2010zd,Fukushima,Kim,Kharzeev,Frolov}, the dense relativistic matter in a strong magnetic 
field may hold some new theoretical surprises. In particular, a topological contribution in the axial current
at the lowest Landau level (LLL) was revealed in Ref.~\cite{Metlitski:2005pr}. More recently,
it was shown in Ref.~\cite{Gorbar:2009bm} that the normal ground state of such matter is characterized by a
chiral shift parameter $\Delta$.} The meaning of this parameter is clearest in the chiral limit: it
determines a relative shift of the longitudinal momenta in the dispersion relations of opposite
chirality fermions, $k^{3}\to k^{3}\pm\Delta$, where the momentum $k^{3}$ is directed along 
magnetic field. Taking into account that fermions in {\em all}
Landau levels, including those around the Fermi surface, are affected by $\Delta$, 
the corresponding matter may have unusual transport and/or emission properties. 

To further justify the motivation for this study, it is instructive to discuss the symmetry properties
of the chiral shift parameter $\Delta$. As we shall see below, it enters the effective Lagrangian
density through the following quadratic term: $\Delta \bar\psi \gamma^3 \gamma^5 \psi $. 
Therefore, just like the external magnetic field, the $\Delta$ term, being symmetric with 
respect to parity transformations ${\cal P}$ , breaks time reversal ${\cal T}$ and the rotational 
symmetry $SO(3)$ down to $SO(2)$ (i.e., the rotations about the axis set by the magnetic field). 
Also, since the $\Delta$ term is even under charge conjugation ${\cal C}$, it breaks ${\cal CPT}$ 
symmetry, which is also broken by the fermion density. We then conclude that the absence of the 
chiral shift parameter is not protected by any symmetry, which, in turn, suggests that such a term 
should be dynamically generated even by perturbative dynamics. It is one of the purposes of 
this paper to shed light on this issue.

The special role of the chiral shift parameter $\Delta$ will be discussed in detail below. 
Already here, however, we would like to point out that the quadratic part of the Lagrangian 
density $\Delta \bar\psi \gamma^3 \gamma^5 \psi $
suggests a possible connection between the parameter $\Delta$ and the axial current along the direction
of the magnetic field. Indeed, the parameter $\Delta$ enters the effective action as a Lagrange multiplier
in front of the operator of the axial current $j^{3}_{5} =  \bar\psi \gamma^3 \gamma^5 \psi$. 
(This could be compared with the role of the Dirac mass, which
formally is the Lagrange multiplier in front of the operator whose ground expectation value is the chiral
condensate.) When the axial current is nonzero in the ground state, it should be generally expected that
$\Delta$ is also nonzero. Now, the axial current is known to be nonzero already in the system of
noninteracting fermions in an external magnetic field \cite{Metlitski:2005pr}. Thus, we argued
in Ref.~\cite{Gorbar:2009bm} that a
nonzero $\Delta$ is an unavoidable consequence in interacting systems and, moreover, it is linear in
the coupling constant to leading order \cite{Gorbar:2009bm}. 
{In this paper we will confirm this suggestion.}

As was pointed out in Ref.~\cite{Metlitski:2005pr}, the structure of the topological axial current, 
induced only in the LLL, is intimately connected with the axial anomaly \cite{ABJ}. This fact is 
directly connected with 
the well known result that in a magnetic field the axial anomaly is also generated only in the LLL 
\cite{Ambjorn}. The important question is whether the form of the induced axial $j^{3}_{5}$
current coincides with the result in the theory of noninteracting fermions in a magnetic field
\cite{Metlitski:2005pr} or whether it is affected by interactions (for related discussions, see
Refs.~\cite{Metlitski:2005pr,Gorbar:2009bm,Rebhan,Fukushima,Rubakov,Hong}). 
As has been recently shown in Ref.~\cite{Gorbar:2010}, while the dynamics responsible for the
generation of the chiral shift $\Delta$ essentially modifies the form of this current, it does {\it not}
affect the form of the axial anomaly. Moreover, while the topological contribution in the axial
current is generated in the infrared kinematic region (at the LLL), the contribution of $\Delta$
in this current is mostly generated in ultraviolet, which implies that higher Landau levels
are important in that case.

The main goal of this paper is to study in detail the dynamics responsible for the generation of 
the chiral shift parameter. This will be done at nonzero temperature and beyond the chiral limit 
in the Nambu-Jona-Lasinio (NJL) model. We also study some general and subtle features of the dynamics with the
chiral shift parameter $\Delta$. In particular it will be shown directly from the {\it form} of the gap 
equation in the NJL model that $\Delta$ necessarily exists in the normal phase in a magnetic field. 
Another property of $\Delta$ that is important for potential applications is that it is rather insensitive 
to temperature when $T \ll \mu_0$, where $\mu_0$ is the chemical potential, and {\it increases} with $T$ 
when $T > \mu_0$. The first regime is appropriate for stellar matter, and the second one is realized in heavy 
ion collisions.
 
Since the NJL model is nonrenormalizable, it is necessary to use a regularization with an ultraviolet
cutoff. The important issue in such a model is how strongly the observables depend on the choice of
a regularization scheme. In Ref.~\cite{Gorbar:2009bm}, a gauge noninvariant regularization (with a cutoff in a sum 
over Landau levels) was used. In this paper, besides that regularization, we will also utilize the gauge invariant 
proper time regularization \cite{Schwinger}. It will be shown that the results in these two
regularization schemes are qualitatively the same.

The rest of this paper is organized as follows. In Sec. \ref{model}, the model of relativistic matter in a magnetic 
field is introduced and  the symmetry properties of various possible dynamical order parameters are overviewed. 
In Sec.~\ref{GapEquation}, we derive
the general form of the Schwinger-Dyson (gap) equation and discuss the approximations used in the analysis
of the dynamics responsible for the chiral symmetry breaking and the generation of the chiral shift parameter.
The analytical solutions to the gap equation at zero temperature are described in Sec. \ref{gapT=0}.
The numerical solutions of the gap equation are presented in Sec.~\ref{Numerical}. Both zero and nonzero
temperature cases are analyzed in detail. The free energies of the corresponding solutions are calculated
and the ground states for various sets of parameters are determined. In Sec.~\ref{sec-axial-current}, the
induced axial current density in this model is calculated and analyzed. In Sec.~\ref{Summary}, we discuss
our main results and their possible applications to the physics of compact stars and heavy ion collisions.
Several Appendices at the end of the paper give many technical details and derivations used in the main text.

\section{Model}
\label{model}

In this paper, in order to reveal the key elements of the dynamics responsible for the generation
of the chiral shift parameter in the clearest way, we use the simplest Nambu-Jona-Lasinio model 
with one fermion flavor. Despite the obvious limitations, such a model with a short-range interaction
is expected to provide a reasonable framework for revealing the qualitative features of the much 
more complicated dynamics in dense QED or QCD plasmas, where the long-range interactions 
are only partially screened.

The Lagrangian density of the model reads
\begin{eqnarray}
{\cal L} &=& \bar\psi \left(iD_\nu+\mu_0\delta_{\nu}^{0}\right)\gamma^\nu \psi
-m_{0}\bar\psi \psi \nonumber\\
&&+\frac{G_{\rm int}}{2}\left[\left(\bar\psi \psi\right)^2
+\left(\bar\psi i\gamma^5\psi\right)^2\right],
\label{NJLmodel}
\end{eqnarray}
where $m_{0}$ is the bare fermion mass and $\mu_0$ is the chemical potential. By
definition, $\gamma^5\equiv i\gamma^0\gamma^1\gamma^2\gamma^3$. The covariant
derivative $D_{\nu}=\partial_\nu -i e A_{\nu}$ includes the external gauge field
$A_{\nu}$. In the presence of a constant magnetic field pointing in the $z$-direction,
the $(3+1)$-dimensional Lorentz symmetry in the model is explicitly broken down
to the $SO(2)$ symmetry of rotations around the $z$-axis in the presence of this magnetic field.
Also, except parity ${\cal P}$, all the discrete symmetries ${\cal C}$,  ${\cal T}$,
${\cal CP}$, ${\cal CT}$, $PT$, and ${\cal CPT}$ are broken.

In the chiral limit, $m_{0}=0$, this model possesses the chiral $U(1)_L\times U(1)_R$
symmetry. In the vacuum state ($\mu_0=0$), however, this chiral symmetry is known to be
spontaneously broken at any $G_{\rm int}>0$ because of the magnetic catalysis phenomenon 
\cite{MC1,MC2}. (For lattice studies of this phenomenon, see Ref.~\cite{Buividovich:2008wf}.) 
In essence, such spontaneous breaking results from the enhanced pairing dynamics of fermions 
and antifermions in the infrared. The enhancement results from the nonvanishing density of
states in the LLL that is subject to an effective dimensional reduction
$D\to D-2$. (This is somewhat reminiscent of the pairing dynamics at the Fermi surface
of a degenerate electron gas in the Bardeen-Cooper-Schrieffer theory of superconductivity.) 
At a sufficiently large value of the chemical potential, the chiral symmetry is expected to be
restored. As we shall see below, this is indeed the case, but the corresponding normal
ground state is characterized by a nonzero chiral shift parameter $\Delta$.

\section{Gap equation at weak coupling}
\label{GapEquation}

\subsection{Structure of the gap equation}
\label{GapA}

As follows from the structure of the Lagrangian density in Eq.~(\ref{NJLmodel}), the
tree level fermion propagator in coordinate space is determined by
\begin{eqnarray}
iS^{-1}(u,u^\prime) &=&\Big[(i\partial_t+\mu_0)\gamma^0
-(\bm{\pi}_{\perp}\cdot\bm{\gamma})\nonumber\\
&&-\pi^{3}\gamma^3-m_{0}\Big]\delta^{4}(u- u^\prime),
\label{sinverse}
\end{eqnarray}
where $u=(t,\mathbf{r})$, while $\pi_{\perp}^{k} \equiv i \partial^k + e A^k$, with $k=1,2$,
and $\pi^{3} = i \partial^3 =- i \partial_3$ are the canonical momenta \cite{footnote-charge}. (Note that the 
components of the conventional gradient $\bm{\nabla}$ are given by $\partial_k\equiv -\partial^k $
and the components of the vector potential $\mathbf{A}$ are identified with the {\em contravariant} 
components of the vector potential $A^k$.) In the rest of this paper, we use the vector potential 
in the Landau gauge, $\mathbf{A}= (0, x B,0)$, where $B$ is the strength of the external magnetic 
field pointing in the $z$-direction.

As for the structure of the full fermion propagator, it is given by the following ansatz:
\begin{eqnarray}
iG^{-1}(u,u^\prime) &=&\Big[(i\partial_t+\mu)\gamma^0 -
(\bm{\pi}_{\perp}\cdot\bm{\gamma})-\pi^{3}\gamma^3
\nonumber\\
&+& i\tilde{\mu}\gamma^1\gamma^2
+\Delta\gamma^3\gamma^5
-m\Big]\delta^{4}(u- u^\prime).
\label{ginverse}
\end{eqnarray}
This propagator contains two new types of dynamical parameters that are absent
at tree level in Eq.~(\ref{sinverse}): $\tilde{\mu}$ and $\Delta$. From its
Dirac structure, it should be clear that $\tilde{\mu}$ plays the role of an anomalous
magnetic moment. As for $\Delta$, it is the chiral shift parameter already mentioned
in the Introduction. Note that in 2 + 1 dimensions (without $z$ coordinate),
$\Delta\gamma^3\gamma^5$ would be a mass term that is odd under time reversal.
This mass is responsible for inducing the Chern-Simons term in the effective action for gauge 
fields \cite{CS}, and it plays an important role in the quantum Hall effect in
graphene \cite{GGM2007,GGMS2008}.

It should be emphasized that the Dirac mass and the chemical potential terms in
the full propagator are determined by $m$ and $\mu$ that may differ
from their tree level counterparts, $m_0$ and $\mu_0$. While $m_0$ is the bare fermion
mass, $m$ has the physical meaning of a dynamical mass that, in general, depends on
the density and temperature of the matter, as well as on the strength of interaction.
Concerning the chemical potentials, it is $\mu_0$ that is the chemical potential in the
thermodynamic sense. The value of $\mu$, on the other hand, is an ``effective" chemical
potential that determines the quasiparticle dispersion relations in interacting theory.

\begin{figure}[ht]
\begin{center}
\includegraphics[width=.45\textwidth]{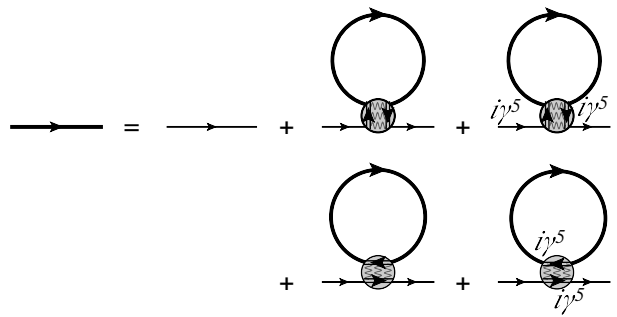}
\caption{Diagrammatic form of the gap equation in the Hartree-Fock 
(mean-field) approximation.} 
\label{fig-sd-eq}
\end{center}
\end{figure}

In order to determine the values of the parameters $m$, $\mu$, $\Delta$ and $\tilde{\mu}$
in the model at hand, we will use the Schwinger-Dyson (gap) equation for the full
fermion propagator. As described in Appendix \ref{AppGapEq}, utilizing the approach based on 
the effective action for composite operators \cite{BK,potential},
one can show that in the mean-field approximation it takes the following form:
\begin{eqnarray}
G^{-1}(u,u^\prime) = S^{-1}(u,u^\prime)
- i G_{\rm int} \left\{ G(u,u) -  \gamma^5 G(u,u) \gamma^5 \right. &&\nonumber \\
- \left. \mbox{tr}[G(u,u)] +  \gamma^5\, \mbox{tr}[\gamma^5G(u,u)]\right\}
\delta^{4}(u- u^\prime). &&\nonumber\\
\label{gap}
\end{eqnarray}
The diagrammatic form of the gap equation is shown in Fig.~\ref{fig-sd-eq}. While 
the first two terms in the curly brackets describe the exchange (Fock) interaction,
the last two terms describe the direct (Hartree) interaction.

This matrix equation is derived in the mean-field approximation,
which is reliable in the weakly coupled regime when the dimensionless coupling constant
\begin{equation}
g \equiv \frac{G_{\rm int}\Lambda^2}{4\pi^2}
\label{g}
\end{equation}
is small, $g \ll 1$.
Here $\Lambda$ is an ultraviolet cutoff, and
the coupling $g$ is defined in such a way that $g_{cr} =1$, where $g_{cr}$ is
the critical value for generating a fermion dynamical mass in the NJL model without
magnetic field.

Of course, weak coupling is completely adequate for the analysis of the electron gas
in the interior of neutron stars. As for the stellar quark matter, such an approximation may
at best provide only a qualitative description of the dynamics responsible for the chiral
asymmetry in the ground state. Regarding the use of the mean-field approximation, there is
no reason to doubt that it should capture the main features of the dynamics, especially in
the weakly coupled limit.

As one can see, the right hand side of the gap equation (\ref{gap}) depends only on the
full fermion propagator $G(u,u^\prime)$ at $u^\prime=u$. This fact greatly simplifies
the analysis. Of course, it is related to the fact that we use the local four-fermion
interaction. This feature will be lost in more realistic models with long-range interactions.

The main disadvantage of the local four-fermion interaction is a nonrenormalizability of
the model. Therefore, the model in Eq.~(\ref{NJLmodel}) should be viewed only as a low-energy effective
model reliable at the energy scales below a certain cutoff energy $\Lambda$. One may try
to associate the value of the cutoff with a certain physical scale, e.g., the Debye
screening mass in dense matter, or another characteristic scale provided by
nonperturbative dynamics. To keep it general, we assume that $\Lambda$ is a free
parameter in the analysis below.

\subsection{Structure of solutions of the gap equation}
\label{GapB}

In this subsection we consider the general structure of the solutions of the gap equation.
In particular, it will be shown directly from the {\it form} of the gap
equation that in the normal phase in a magnetic field
a) $\mu$ and $\mu_0$ are different, and
b) the shift parameter $\Delta$ is necessarily nonzero.
 
As shown in Appendix \ref{AppGapEq}, in the mean-field approximation utilized here, 
the Dirac structure of gap equation (\ref{gap}) does not allow solutions with a nontrivial
$\tilde{\mu}$. While having $\tilde{\mu} = 0$ simplifies the analysis, we should emphasize
that $\tilde{\mu}$ may well be nonvanishing in more refined approximations and in models
with other types of interactions \cite{GGM2007,GGMS2008,magCatFI}. At the same time, as
one learns from a similar analysis in graphene, a nonzero $\tilde{\mu}$ should not change
the main qualitative features of the phase with an induced $\Delta$ \cite{GGM2007,GGMS2008}.

The explicit expression for $G(u,u)$ is calculated in Eq.~(\ref{Guu}) in 
Appendix~\ref{AppPropagator}. The result reads
\begin{equation}
G(u,u) =
\frac{i}{2\pi l^2}\sum_{n=0}^{\infty} \int\frac{d\omega d k^{3}}{(2\pi)^2}
\frac{{\cal K}_{n}^{-}{\cal P}_{-}+{\cal K}_{n}^{+}{\cal P}_{+}\theta(n-1)}{U_n},
\label{full-propagator}
\end{equation}
where $l=1/\sqrt{|e B |}$ is the magnetic length,
$\theta(n-1)\equiv 1$ for $n\geq 1$ and $\theta(n-1)\equiv0$ for $n\leq 0$.
We also use the following spin
projectors:
\begin{equation}
{\cal P}_{\pm} =\frac12\left(1\pm is_{\perp}\gamma^1\gamma^2\right),
\label{projectorP}
\end{equation}
and the shorthand notation $s_{\perp}\equiv \sign(e B )$. The functions $U_n$ and
${\cal K}_{n}^{\pm}$ are defined in Eqs.~(\ref{U_n}) and (\ref{K_n^pm}), respectively.
For reader's convenience, here they are quoted only for the case of vanishing 
$\tilde\mu$, which is of the main interest:
\begin{widetext}
\begin{eqnarray}
{\cal K}_{n}^{\pm}&=& \left[(\omega+\mu \mp s_{\perp}\Delta)\gamma^0
+ m  -k^{3}\gamma^3\right]\left[
(\omega+\mu)^2  - m^2 - \Delta^2 - (k^{3})^2 - 2n|e B |
\mp 2 s_{\perp}\Delta \left(m + k^{3} \gamma^3 \right)\gamma^0
\right],
\label{K_n^pm_text}
\\
U_n &=&
 \left[(\omega+\mu)^2 - 2n|e B | -\left(s_\perp \Delta-\sqrt{m^2+(k^{3})^2}\right)^2\right]
 \left[(\omega+\mu)^2 - 2n|e B | -\left(s_\perp \Delta+\sqrt{m^2+(k^{3})^2}\right)^2\right].
\label{U_n_text}
\end{eqnarray}
By making use of these expression, the $n$th Landau level contribution to the
fermion propagator can be cast in the following form:
\begin{equation}
\frac{{\cal K}_{n}^{\pm} {\cal P}^{\pm}}{U_n} =  \gamma^0 \left[
\frac{\omega+\mu -i s_{\perp} \gamma^1\gamma^2 \left(s_{\perp} \Delta - \sqrt{m^2+(k^{3})^2} \right)}
{(\omega+\mu)^2 - \left(s_{\perp}\Delta-\sqrt{m^2+(k^{3})^2} \right)^2 - 2n|e B |}{\cal H}^{-}
+
\frac{\omega+\mu -i s_{\perp} \gamma^1\gamma^2 \left(s_{\perp} \Delta + \sqrt{m^2+(k^{3})^2} \right)}
{(\omega+\mu)^2 - \left(s_{\perp}\Delta+\sqrt{m^2+(k^{3})^2} \right)^2 - 2n|e B |}{\cal H}^{+}
\right]{\cal P}^{\pm},
\label{K_P_U}
\end{equation}
\end{widetext}
where
\begin{equation}
{\cal H}^{\pm}=\frac{1}{2}\left(1 \pm s_{\perp}\frac{ m  +k^{3}\gamma^3}{\sqrt{m^2+(k^{3})^2} }\gamma^5\gamma^3\right)
\label{projectorH}
\end{equation}
are the projectors on the quasiparticle states, whose energies are given in terms of either the sum 
or the difference of $s_{\perp}\Delta$ and $\sqrt{m^2+(k^{3})^2}$. The projectors take a particularly simple form 
in the massless limit: ${\cal H}_{m=0}^{\pm}=\frac{1}{2}\left[1 \pm s_{\perp}\sign(k^3)\gamma^5 \right]$. 
In this case, ${\cal H}_{m=0}^{\pm}$ almost coincide (up to the sign of the longitudinal momentum) 
with the chirality projectors ${\cal P}_{5}^{\pm}=\frac{1}{2}\left(1 \pm s_{\perp} \gamma^5 \right)$, used 
in Ref.~\cite{Gorbar:2009bm}. For each choice of the signs of $eB$ and $k^3$, the chirality of the states 
that correspond to projectors ${\cal H}^{\pm}_{m=0}$ are summarized in Table~\ref{tab-I}. 
In fact, the precise relation between the two sets of projectors reads
\begin{equation}
{\cal H}_{m=0}^{\pm}=\frac{1 \mp \sign(k^3)}{2} {\cal P}_{5}^{-}+\frac{1 \pm \sign(k^3)}{2} {\cal P}_{5}^{+}.
\label{Hpm-P5pm}
\end{equation}
By making use of this relation and taking into account that $ \sqrt{m^2+(k^{3})^2}\to |k^{3}|$ when $m\to0$, 
it is straightforward to check that the propagator in Eq.~(\ref{full-propagator}) in the massless limit takes 
exactly the same form as in Ref.~\cite{Gorbar:2009bm}, i.e.,  
\begin{equation}
G(u,u)=G_0^{-}{\cal P}_{-}
+\sum_{n=1}^{\infty}\left(G_{n}^{-}{\cal P}_{-}+G_{n}^{+}{\cal P}_{+}\right),
\end{equation}
where 
\begin{eqnarray}
G_{n}^{\pm} &=& 
\frac{i|eB|\gamma^0}{2\pi}\int\frac{d\omega d k^3}{(2\pi)^2} \nonumber\\
&&\times \Big[
 \frac{\omega+\mu \pm[k^3-\Delta \sign(eB)]}
      {(\omega+\mu)^2 - 2n|eB| - [k^3- \Delta\sign(eB)]^2 }{\cal P}_{5}^{-}
\nonumber\\
&&+\frac{\omega+\mu \mp[k^3+\Delta \sign(eB)]}
      {(\omega+\mu)^2 - 2n|eB| - [k^3+ \Delta\sign(eB)]^2 }{\cal P}_{5}^{+}
\Big].\nonumber\\
\end{eqnarray}   
The opposite chirality fermions, described by such a propagator, are characterized by a relative 
shift of the longitudinal momenta, $k^{3}\to k^{3}\pm s_{\perp}\Delta$, in their dispersion relations.

\begin{table}
\caption{Chirality of the eigenstates that correspond to projectors ${\cal H}_{m=0}^{\pm}$ for each sign of 
the longitudinal momentum $k^3$.}
\label{TableI}
\begin{ruledtabular}
\begin{tabular}{ c | cc cc}
       & \multicolumn{2}{c}{ ${\cal H}_{m=0}^{-}$} &  \multicolumn{2}{c}{${\cal H}_{m=0}^{+}$}\\
       & $k^3<0$  & $k^3>0$  & $k^3<0$  & $k^3>0$ \\
\hline
$\sign(eB)>0$ & L &  R  &  R  & L  \\
$\sign(eB)<0$ & R  & L  &  L  &  R \\
\end{tabular}
\end{ruledtabular}
\label{tab-I}
\end{table} 

Unlike the higher Landau level terms in the propagator, the LLL contribution is rather simple,
\begin{eqnarray}
\frac{{\cal K}_{0}^{-} {\cal P}^{-}}{U_0} 
&=&\gamma^0  \left[\frac{{\cal H}^{-}}
{\omega+\mu- s_{\perp} \Delta + \sqrt{m^2+(k^{3})^2} } \right. \nonumber\\
&+&\left.  \frac{{\cal H}^{+}}
{\omega+\mu - s_{\perp} \Delta - \sqrt{m^2+(k^{3})^2} }\right]{\cal P}^{-}.
\label{K_P_U-LLL}
\end{eqnarray}
In the LLL, $s_{\perp} \Delta$ is a part of the effective chemical potential $\mu - s_{\perp} \Delta$,
and the two terms in Eq.~(\ref{K_P_U-LLL}) can be associated with the antiparticle (negative energy) and 
particle (positive energy) contributions, respectively. In order to avoid a potential confusion, let us also mention 
that, as seen from Eq.~(\ref{K_P_U}), the connection between ${\cal H}^{\pm}$ and the particle/antiparticle 
states is not preserved in the higher Landau levels. 

As follows from Eq.~(\ref{K_P_U}) and Eq.~(\ref{K_P_U-LLL}), the poles of the full fermion 
propagator are at
\begin{equation}
\omega_0 = -\mu + s_{\perp} \Delta\pm\sqrt{m^2+(k^{3})^2},
\label{omega0}
\end{equation}
for the lowest Landau level, and at
\begin{equation}
\omega_n = -\mu\pm \sqrt{\left(s_{\perp}\Delta\pm\sqrt{m^2+(k^{3})^2}\right)^2 + 2n|e B | },
\label{omegan}
\end{equation}
for higher Landau levels, $n\geq 1$. Note that all four combinations of signs are possible for
the latter.

The general form of the gap equation at nonzero temperature is derived in Appendix~\ref{AppGapEq}, see
Eqs.~(\ref{gap-mu}), (\ref{gap-Delta}) and (\ref{gap-m}),
\begin{eqnarray}
\mu &=&\mu_0 -\frac{1}{2}G_{\rm int}  {\cal A}  ,
\label{gap-mu-text}  \\
m &=& m_0 - G_{\rm int}  {\cal B}    ,
\label{gap-m-text}\\
\Delta &=& -\frac{1}{2}G_{\rm int} {\cal D}    .
\label{gap-Delta-text} 
\end{eqnarray}
The functions ${\cal A}$, ${\cal B}$ and ${\cal D}$ on the right hand side of these equations are
determined by the full fermion propagator as follows:
\begin{eqnarray}
 {\cal A} &=&-\tr\left[\gamma^0G(u,u)\right] \equiv \langle j^{0}\rangle , 
 \label{A-density-text}\\
{\cal B}&=& -\tr\left[G(u,u)\right] \equiv \langle \bar{\psi}\psi\rangle ,
\label{B-chi-condensate-text}\\
{\cal D} &=&-\tr\left[\gamma^3\gamma^5G(u,u)\right] \equiv \langle j^{3}_5\rangle ,
\label{D-axial-current-text}
\end{eqnarray}
and have the meaning of the fermion charge density, the chiral condensate and the axial current 
density, respectively.
The formal integral representations of these three functions are presented in Eqs.~(\ref{functionA}),  
(\ref{functionB}), and (\ref{functionD}). Two of them, ${\cal B}$ and ${\cal D}$, contain ultraviolet 
divergences. In the next section, we study  various solutions to the gap equations by using two 
regularization schemes: a gauge noninvariant one, with cutoffs in momentum integration and 
the sum over the Landau levels \cite{Gorbar:2009bm}, and the gauge invariant proper-time 
regularization \cite{Schwinger}.

Let us now consider the zero temperature normal phase in the chiral limit, when $m=m_0=0$
and $\langle \bar\psi \psi\rangle = 0$. It is realized when the chemical potential
$\mu_0 > m_{dyn}/\sqrt{2}$ (see Ref.~\cite{Gorbar:2009bm} and Sec. \ref{gapT=0} below), 
where $m_{dyn}$ is a dynamical
fermion mass in a magnetic field at zero chemical potential and zero temperature. Let us
analyze Eqs.~(\ref{gap-mu-text}) and (\ref{gap-Delta-text}) in perturbation theory in the 
dimensionless coupling constant $g$ defined in Eq.~(\ref{g}).
In the zero order approximation, we have a theory of free fermions in a magnetic
field. To this order, $\mu=\mu_0$ and $\Delta=0$. However, even in this case the fermion
density $\langle j^0\rangle$ and the axial current density $\langle j_5^3\rangle$ are
nonzero. The former can be presented as a sum over the Landau levels:
\begin{eqnarray}
\langle j^0\rangle_0 &=& \frac{ \mu_0|eB|}{2\pi^2} +\frac{\sign(\mu_0)|eB|}{\pi^2}
\sum_{n=1}^{\infty}
\sqrt{\mu_{0}^2-2n|eB|}\nonumber\\
&&\times \theta\left(|\mu_0|-\sqrt{2n|eB|}\right),
\label{j}
\end{eqnarray}
and the latter comes entirely from the LLL \cite{Metlitski:2005pr}:
\begin{equation}
\langle j^3_5\rangle_0 = \frac{-eB}{2\pi^2}\mu_0\, .
\label{MZ}
\end{equation}
(The overall minus sign is due to our convention for the electric charge of the electron 
\cite{footnote-charge}.) 
Then, to the next order in the coupling constant, one finds from 
Eq.~(\ref{gap-Delta-text}) that $\Delta \propto G_{\rm int} \langle j^3_5\rangle_0 \neq 0$. 
Thus, in the normal phase of this theory, there 
{\it necessarily} exists a shift parameter $\Delta$.
In essence, the latter is one of the main results of Ref.~\cite{Gorbar:2009bm}.
Let us also emphasize that $\Delta$ is generated by perturbative dynamics, which is directly
connected with the fact that the vanishing $\Delta$ is not protected by any symmetry (recall
that ${\cal C} =+1$, ${\cal P} =+1$, and ${\cal T} = -1$ for the axial current density $j_5^3$,
and beside parity ${\cal P}$, all the discrete symmetries are broken in model (\ref{NJLmodel})).

This result was obtained for the case of zero temperature. As will be shown below in 
Sec.~\ref{sec-axial-current}, the chiral shift parameter is rather insensitive to the value 
of the temperature in the regime of cold dense matter appropriate for potential applications in 
stars. In the case of heavy ion collisions, as we shall see,
a temperature larger than the chemical potential may play an 
important role in enhancing the chiral shift parameter.

As one can see from Eqs.~(\ref{gap-mu-text}), (\ref{A-density-text}), and (\ref{j}), $\mu - \mu_0 
\propto G_{\rm int}\langle j^0\rangle_0 \neq 0$, which implies that $\mu$ and $\mu_0$ are different.
The origin of this difference can be traced to the Hartree terms in the gap equation [see the last two
terms in Eq.~(\ref{gap})]. 

This finding seems to be robust in the NJL model with a local four-fermion interaction and a 
chemical potential, associated with a global charge, such as a baryon (or lepton) charge for example.
When the conserved charge is related to a gauge symmetry, as in the case of the electric charge,
the situation may be different. In that case, a neutrality condition imposed by the Gauss law
takes place \cite{Kapusta}. The latter is necessary for providing the thermodynamic equilibrium
in a system. This is likely to result in $\mu^{(e)}=\mu_{0}^{(e)}$ when
$\mu^{(e)}$ is the chemical potential for
electric charge. Note that usually there are chemical potentials of both types in dense
relativistic matter. While being of importance for potential applications in principle, we expect
that this fact will not change our main conclusion regarding the chiral shift parameter. 

In conclusion, let us briefly discuss the following issue. One may think that when
the fermion mass $m$ is zero, the term with the chiral shift $\Delta$ is unphysical: 
in this case, it could formally be removed by the gauge transformation $\psi\to e^{iz\gamma_5 
\Delta}\psi$, $\bar{\psi} \to \bar{\psi}e^{iz\gamma_5\Delta}$.  
The point, however, is that this transformation is 
singular (anomalous). It follows 
from the two facts: (i) as was already pointed out above, in the LLL, $s_{\perp}\Delta$ 
is a part of the chemical potential (see Eq.~(\ref{K_P_U-LLL})),
and (ii) this happens because the LLL dynamics is $1 + 1$-dimensional \cite{MC2}.
It is well known that in $1 + 1$ dimensions this transformation, which formally varies the value of the 
chemical potential, is anomalous (for a recent thorough discussion of this transformation, 
see Ref.~\cite{Kojo}).

\section{Analytical solutions at $T=0$}
\label{gapT=0}

To set up a benchmark for the numerical results, it is instructive to start the analytical analysis 
of gap equations (\ref{gap-mu-text}), (\ref{gap-m-text}), and (\ref{gap-Delta-text}) at zero 
temperature. We will use two regularization schemes: (i) the gauge noninvariant one, with 
a sharp momentum cutoff, $|k^{3}|\leq \Lambda$, in the integrals over $k^{3}$ (which are 
always performed first) and a smooth cutoff in the sums over the Landau levels (which are 
performed last), and (ii) the gauge invariant proper-time regularization. It will be shown that 
the results in these two regularizations are qualitatively the same.

\subsubsection{Analytical solutions in the momentum cutoff regularization}

Let us start from the first regularization. The smoothing function in the sums over the 
Landau levels is taken in the following form:
\begin{equation}
\kappa(n)= \frac{\sinh(\Lambda/\delta\Lambda)}
{\cosh\left((\Lambda/\delta\Lambda)\sqrt{n/n_{\rm cut}}\right)+\cosh(\Lambda/\delta\Lambda)},
\label{kappa}
\end{equation}
where the cutoff value $n_{\rm cut}$ is determined by the number of the Landau levels below the energy
scale set by $\Lambda$, i.e., $n_{\rm cut} \equiv \left[\Lambda^2/2|eB|\right]$ with the square brackets
denoting the integer part. The width of the energy window in which the cutoff is smoothed is determined
by the ratio $\Lambda/\delta\Lambda$, and when the value of the latter goes to infinity, 
the function $\kappa(n)$ approaches the step function. (In the numerical calculations below 
we use $\Lambda/\delta\Lambda=20$.)

Let us now show that there are two qualitatively 
different solutions, which were previously reported in Ref.~\cite{Gorbar:2009bm}.

{\it Solution of Type I}.
The first solution type corresponds to $m\neq 0$ and $\Delta=0$ in accordance with the magnetic
catalysis scenario in the vacuum \cite{MC1,MC2}. By substituting $m\neq 0$ and
$\Delta=0$ into the general expressions (\ref{functionA}), (\ref{functionB}), and (\ref{functionD}),
we derive the following expressions for the functions appearing on the right hand sides of 
Eqs.~(\ref{gap-mu-text})-(\ref{gap-Delta-text}):
\begin{widetext}
\begin{eqnarray}
{\cal A} &=& \frac{\sign(\mu)}{2(\pi l)^2}\sqrt{\mu^2-m^2} \theta\left(|\mu|-|m|\right)
+\frac{\sign(\mu)}{(\pi l)^2}\sum_{n=1}^{\infty} \sqrt{\mu^2-2n|eB|-m^2}\,
\theta\left(|\mu|-\sqrt{m^2+2n|eB|}\right),
\label{AT0}
\\
{\cal B}&\simeq & -\frac{m}{2(\pi l)^2}\left[\ln\frac{2\Lambda}{|m|}
-\ln\frac{|\mu|+\sqrt{\mu^2-m^2}}{|m|}\theta\left(|\mu|-|m|\right) \right]
-\frac{m}{(\pi l)^2}\sum_{n=1}^{\infty} \left[\ln\frac{\Lambda+\sqrt{\Lambda^2+m^2+2n|eB|}}{\sqrt{m^2+2n|eB|}}\right.
\nonumber \\
&&\left.
-\ln\frac{|\mu|+\sqrt{\mu^2-m^2-2n|eB|}}{\sqrt{m^2+2n|eB|}}
\theta\left(|\mu|-\sqrt{m^2+2n|eB|}\right)\right],
\label{BT0}
\\
{\cal D}&\simeq &-s_\perp\frac{\sign(\mu)}{2(\pi l)^2}\sqrt{\mu^2-m^2} \theta\left(|\mu|-|m|\right).
\label{DT0}
\end{eqnarray}
\end{widetext}
Note that the expression for ${\cal D}$ is proportional to the LLL contribution to the fermion density
and, as a result, vanishes when $|\mu|<|m|$. In this case, a solution with $\Delta=0$ is consistent with
gap equation (\ref{gap-Delta-text}). Then, the other two gap equations reduce down to $\mu =\mu_0$
and
\begin{eqnarray}
m &=& m_0 + \frac{2 g m}{(\Lambda l)^2}
\Big[\ln\frac{2\Lambda}{|m|}\nonumber\\
&+& 2\sum_{n=1}^{\infty}\kappa(n)\ln\frac{\Lambda+\sqrt{\Lambda^2+m^2+2n|eB|}}{\sqrt{m^2+2n|eB|}}
\Big],
\label{gap-m-Delta0}
\end{eqnarray}
where we utilized the smooth cutoff function (\ref{kappa}) in the sum over the Landau levels. 

{\it Solution of Type II}.
In the chiral limit, in addition to the solution with a nonzero Dirac mass $m$, the gap equation
also allows a solution with $m=0$ and a nonzero chiral shift parameter $\Delta$. To see this,
we derive the functions that appear on the right hand sides of the gap equations for this
special case:
\begin{eqnarray}
{\cal A}&=& \frac{ \mu-s_{\perp}\Delta }{2(\pi l)^2} +\frac{\sign(\mu)}{(\pi l)^2} \sum_{n=1}^{N_B}
\sqrt{\mu^2-2n|eB|}   ,
\label{AT0m0}
\\
{\cal B} &=& 0 ,
\label{BT0m0}
\\
{\cal D} &=& -\frac{1}{2(\pi l)^2}\left(s_{\perp} \mu - \Delta
-2 \Delta \sum_{n=1}^{\infty}\kappa(n)\right),
\label{DT0m0}
\end{eqnarray}
where $N_B$ is the integer part of $\mu^2/(2|eB|)$.
(Here it might be appropriate to note that the above result for ${\cal D}$ remains
unchanged also at nonzero temperatures!) The fact that
now ${\cal B} = 0$ is in agreement with Eq.~(\ref{gap-m-text})
and the assumption $m=m_0=0$. The remaining two equations, (\ref{gap-mu-text}) and (\ref{gap-Delta-text}),
reduce down to
\begin{equation}
\mu =\mu_0+  \frac{g}{(\Lambda l)^2}\left( s_{\perp}\Delta -\mu -2 \sign(\mu)\sum_{n=1}^{N_B}\sqrt{\mu^2-2n|eB|}\right)
\label{TypeII-mu-equation}
\end{equation}
and
\begin{equation}
\Delta = \frac{g }{(\Lambda l)^2}\left(s_{\perp} \mu - \Delta
-2 \Delta \sum_{n=1}^{\infty}\kappa(n)\right),
\label{TypeII-Delta-equation}
\end{equation}
respectively. To leading order in the coupling constant, the solutions for $\mu$ and $\Delta$
are straightforward,
\begin{eqnarray}
\mu & \simeq & \frac{\mu_0}{1+g/(\Lambda l)^2},   
\label{mu-cutoff}\\
\Delta &=& \frac{g s_{\perp} \mu}{(\Lambda l)^2 +g \left[1 +2\sum_{n=1}^{\infty}\kappa(n)\right]},
\label{Delta-cutoff}
\end{eqnarray}
which are derived under the assumption that the chemical potential $\mu_0$ is not large enough for the
first Landau level to start filling up, i.e., $|\mu|\lesssim \sqrt{2 |eB|}$. When the chemical potential
becomes larger, the result for $\mu$ will get corrections, but the expression for $\Delta$ in terms of
$\mu$ will keep the same form. 
Note that with the function $\kappa(n)$ given in Eq.~(\ref{kappa}), one 
finds that 
\begin{equation}
\sum_{n=1}^{\infty} \kappa(\sqrt{2n|eB|},\Lambda) = a \Lambda^2/|eB|\, ,
\label{sumkappa}
\end{equation}
where $a=O(1)$.

\subsubsection{Analytical solutions in the proper-time regularization}

{\it Solution of Type I}. 
In the regime of magnetic catalysis, we have shown above that the dynamical mass parameter satisfies 
Eq.~(\ref{gap-m-Delta0}) in the momentum cutoff regularization scheme. Now, let us 
show that this is consistent with 
the result obtained in the proper-time regularization. The expression for the vacuum part of function 
${\cal B}$ in this regularization
is given in Eq.~(\ref{B-vacuum}). To the leading logarithm order, then, we derive the following 
gap equation for the mass parameter:
\begin{equation}
m=m_0 +\frac{mg}{(\Lambda l)^2}\ln\frac{1}{\pi(ml)^2}.
\label{gap-proper-time}
\end{equation}
As is easy to check, this gap equation is equivalent to Eq.~(\ref{gap-m-Delta0})  to leading order. 
The corresponding  gap equation is also in agreement with the proper-time result in 
Ref.~\cite{MC2}, which was given in the form
\begin{equation}
m = m_0 + \frac{g m}{(\Lambda l)^2} \int_{1/\Lambda^2}^{\infty}\frac{ds}{s} e^{-s m^2}
\coth\left(|eB|s\right),
\label{gap-m-propertime}
\end{equation}
where the proper-time cutoff $s_0\equiv 1/\Lambda^2$ is
conventionally given in terms of an ultraviolet energy scale $\Lambda$.
By noting that $\coth\left(|eB|s\right)\equiv 1+2\sum_{n=1}^{\infty}e^{-2sn|eB|}$ and using the
table integral
\begin{equation}
\int_{1/\Lambda^2}^{\infty}\frac{ds}{s} e^{-s a^2}
=\Gamma(0,\frac{a^2}{\Lambda^2})\simeq \ln\frac{\Lambda^2}{a^2}+O(1),
\end{equation}
we also see that, up to higher order corrections in powers of $1/\Lambda^2$, the representation 
in Eq.~(\ref{gap-m-propertime}) is equivalent to Eq.~(\ref{gap-m-Delta0}).

As follows from Eq.~(\ref{gap-proper-time}), the value of the dynamical Dirac mass $m_{\rm dyn}$ 
in the chiral limit, $m_0=0$, reads \cite{MC2},
\begin{equation}
m^{2}_{\rm dyn} =\frac{1}{\pi l^2}\exp\left(-\frac{(\Lambda l)^2}{g}\right).
\label{DiracMass}
\end{equation}
Formally, this solution exists for $|\mu_0| < m_{\rm dyn}$. As we will discuss below, however, it
corresponds to the ground state only in a part of this range, $|\mu_0| \lesssim m_{\rm dyn}/\sqrt{2}$.

{\it Solution of Type II}. 
Now let us consider the chiral limit and search for a solution with $m=0$ and a nonzero chiral shift 
parameter $\Delta$ using the proper-time representation. In this case, ${\cal B}=0$ and the expressions 
for ${\cal A}$ and ${\cal D}$ are derived in Appendix~\ref{AppFunctionsABD}. Function ${\cal A}$ 
is finite and, therefore, is given by the same expression as in Eq.~(\ref{AT0m0}). Function ${\cal D}$ 
contains ultraviolet divergences. The corresponding regularized vacuum contribution is given in 
Eq.~(\ref{D-vacuum}). By adding also the (finite) matter contribution, derived in Eq.~(\ref{D-matter}), 
we obtain
\begin{equation}
{\cal D} = \frac{\sqrt{\pi}\Lambda}{2(2\pi l)^2} e^{-(\Delta/\Lambda)^2 }
\mbox{erfi}\left(\frac{\Delta}{\Lambda} \right)\coth\left(\frac{eB}{\Lambda^2} \right)
-\frac{2 s_{\perp}\mu}{(2\pi l)^2} ,
\end{equation}
where $\mbox{erfi}(x) \equiv -i\mbox{erf}(ix)$ 
is the imaginary error function. By expanding the expression for 
${\cal D}$ in inverse powers of $\Lambda$, we arrive at the following approximate result:
\begin{equation}
{\cal D} \simeq -\frac{1}{2(\pi l)^2}\left( s_{\perp}\mu-\frac{(\Lambda l)^2}{2}\Delta\right),
\label{DT0m0-proper-time}
\end{equation}
which is in agreement with the result in Eq.~(\ref{DT0m0}) after making the identification
$\frac{1}{2}(\Lambda l)^2 \equiv 1 +2\sum_{n=1}^{\infty}\kappa(n) \simeq 2a(\Lambda l)^2$, where
the parameter $a$ is defined in Eq.~(\ref{sumkappa}). As we see, $a= 1/4$ in the proper-time regularization.

The gap equation for $\mu$ is insensitive to the ultraviolet dynamics and coincided with Eq.~(\ref{TypeII-mu-equation}). 
By making use of the approximation in Eq.~(\ref{DT0m0-proper-time}), we arrive at the following 
equation for $\Delta$:
\begin{equation}
\Delta  = \frac{g }{(\Lambda l)^2}\left(s_{\perp} \mu - \frac{(\Lambda l)^2}{2}\Delta \right),
\end{equation}
which is equivalent to Eq.~(\ref{TypeII-Delta-equation}) after the same identification of the regularization schemes 
is made. Also, the proper-time solution,
\begin{eqnarray}
\mu & \simeq & \frac{\mu_0}{1+g/(\Lambda l)^2},\\
\Delta &=& \frac{g s_{\perp} \mu}{(\Lambda l)^2 +\frac{1}{2}g (\Lambda l)^2},
\label{Delta-proper-time}
\end{eqnarray}
is equivalent to the solution in Eqs.~(\ref{mu-cutoff}) and (\ref{Delta-cutoff}).

\subsubsection{Free energy}

As should be clear from the above discussion, in the region $|\mu_0|< m_{\rm dyn}$, the two
inequivalent solutions coexist. In order to decide which of them describes the ground state, one
has to compare the corresponding free energies. The general expression for the free energy density
is derived in Appendix~\ref{free-energy}. For the two cases of interest here, the corresponding
results are given in Eqs.~(\ref{AppAOmega0}) and (\ref{AppAOmegaDelta0}),
\begin{equation}
\Omega_{m} \simeq  -\frac{m_{\rm dyn}^2}{2(2\pi l)^2}
\left(1+(m_{\rm dyn} l)^2\ln|\Lambda l|\right)
\label{Omega0}
\end{equation}
and
\begin{equation}
\Omega_{\Delta} \simeq -\frac{\mu_0^2}{(2\pi l)^2}\left(1-g\frac{|eB|}{\Lambda^2}\right),
\label{OmegaDelta0}
\end{equation}
respectively. In deriving the last expression, we used the approximate relations
$\mu\simeq \mu_0$ and $\Delta \simeq g\mu_0 eB/\Lambda^2$. By comparing the free
energies in Eqs.~(\ref{Omega0}) and (\ref{OmegaDelta0}), we see that the ground
state with a nonzero $\Delta$ becomes favorable when $\mu_0\gtrsim m_{\rm dyn}/\sqrt{2}$.
This is analogous to the Clogston relation in superconductivity \cite{Clogston}.

\section{Numerical solutions to the gap equation}
\label{Numerical}

In order to solve numerically the set of gap equations (\ref{gap-mu-text}), (\ref{gap-m-text}), 
and (\ref{gap-Delta-text}), we have to regulate the divergences that appear in the integrals over the
longitudinal momentum $k^{3}$ and the sums over the Landau levels in the expressions for 
the chiral condensate ${\cal B}$ and the axial current density ${\cal D}$. In Sec. \ref{gapT=0} we used
two regularizations: 1) with a sharp momentum cutoff, $|k^{3}|\leq \Lambda$, in
the integrals over $k^{3}$ (which are always performed first) and a smooth cutoff in the sums
over the Landau levels, and 2) the proper-time regularizations. Because it was shown that at zero temperature 
the results in these two regularizations are qualitatively similar, we perform a detailed numerical analysis of the 
gap equations at arbitrary temperature by using the first regularization only, which is technically much simpler to 
implement.

The form of the smoothing function $\kappa(n)$ in this regularization is given in Eq.~(\ref{kappa}).
The width of the energy window in which the cutoff is smoothed is determined
by the ratio $\Lambda/\delta\Lambda$. When the value of this ratio goes to infinity, $\kappa(n)$
approaches a step function, $\theta(n_{\rm cut}-n)$, corresponding to the case of a sharp cutoff at
$n_{\rm cut} = \left[\Lambda^2/2|eB|\right]$. We note, however, that taking a very sharp cutoff in the sums 
over the Landau levels may
result in some unphysical discontinuities in the physical properties of the model as a function of the
magnetic field. This is because of the discontinuities in the dependence of the function $n_{\rm cut}(|eB|)$,
which defines the number of the dynamically accessible Landau levels. In our numerical calculations,
we choose a reasonably large value $\Lambda/\delta\Lambda =20$.

In order to keep our model study as general as possible, we specify all energy/mass parameters in units
of the cutoff parameter $\Lambda$. In the numerical calculations below, we use the following values of
the coupling constant and the magnetic field:
\begin{eqnarray}
g &=& \frac{G_{\rm int}\Lambda^2}{4\pi^2} =0.25,
\label{g-value}\\
|eB| &=& 0.125\Lambda^2 .
\label{eB-value}
\end{eqnarray}
The coupling is rather weak to justify the approximations used in the analysis. In real dense
or hot quark matter, the actual value of dimensionless coupling may be even stronger. In the
degenerate electron gas in the interior of compact stars, on the other hand, it is still much
weaker. Our purpose here, however, is to perform a qualitative analysis of the model and
reveal the general features of the dynamics relevant for the generation of the chiral shift
parameter. Therefore, our ``optimal" choice of $g$ is sufficiently weak to make the analysis
reliable, while not too weak to avoid a very large hierarchy of the energy scales which would
make the numerical analysis too difficult. Similar reasoning applies to the choice of the
magnetic field in Eq.~(\ref{eB-value}). This is a sufficiently strong field that makes it
easier to explore and understand the qualitative features of the dynamics behind both the
magnetic catalysis and the generation of the chiral shift parameter. In applications related
to compact stars, the actual fields might be considerably weaker. However, this value may in
fact be reasonable for applications in heavy ion collisions \cite{Kharzeev:2007tn,Skokov:2009qp}.

\subsection{Numerical solutions at $\mu_0=0$}
\label{gapMu=0numerical}

The gap equation is solved by multiple iterations of the gap equations. The
convergence is checked by measuring the following error function:
\begin{equation}
\epsilon_n =\sqrt{\sum_{i=1}^{3}\frac{(x_{i,n}-x_{i,n-1})^2}{\mbox{max}(x_{i,n}^2,x_{i,n-1}^2)}},
\end{equation}
where ${x}_i=\mu,\Delta,m$ for $i=1,2,3$, respectively. (In the case when both $x_{i,n}$
and $x_{i,n-1}$ vanish, the corresponding $i$th contribution to $\epsilon_n$ is left out.)
When the value of $\epsilon_n$ becomes less than $10^{-4}$ (at $T\neq 0$) or $10^{-5}$
(at $T=0$), the current set of $\mu_{n}$, $\Delta_{n}$ and $m_{n}$ is accepted as an
approximate solution to the gap equation. Usually, the convergence is achieved after
several dozens of iterations. In some cases, even as few as five iterations suffices
to reach the solution with the needed accuracy. This is often the case when we
automatically sweep over a range of values of some parameter (e.g., the temperature
or the chemical potential) and use the solution obtained at the previous value of the
parameter as the starting guess to solve the equation for a new nearby value of the same
parameter. However, even in this approach, the required number of iterations may
sometimes be in the range of hundreds. This is usually the case when the dynamically
generated $\Delta$ and $m$ have a steep dependence on the model parameters,
which is common, e.g., in the vicinity of phase transitions.

\begin{figure}[ht]
\begin{center}
\includegraphics[width=.45\textwidth]{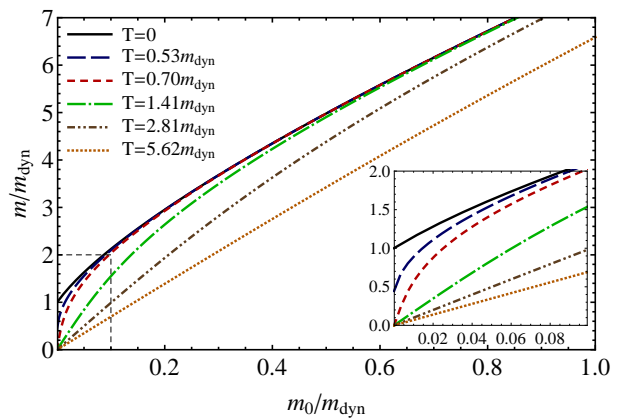}
\caption{(Color online) The dependence of the constituent mass $m$ on the bare
mass $m_0$  at $\mu_0=0$ for several fixed values of temperature:
$T=0$ (black), 
$T=3.75\times 10^{-4}\Lambda=0.53m_{\rm dyn}$ (blue), 
$T=5\times 10^{-4}\Lambda=0.70m_{\rm dyn}$ (red), 
$T=10^{-3}\Lambda=1.41m_{\rm dyn}$ (green),
$T=2\times 10^{-3}\Lambda=2.81m_{\rm dyn}$ (dark brown), and 
$T=4\times 10^{-3}\Lambda=5.62m_{\rm dyn}$ (light brown). 
The insert shows the details in the rectangular area around the origin.} 
\label{figMass-vs-m0}
\end{center}
\end{figure}

In order to set up the reference point for the nonzero chemical potential calculations,
let us start by presenting the results for the constituent fermion mass as a function
of the bare mass $m_0$ at $\mu_0=0$. The corresponding zero temperature dependence is
shown by the black line in Fig.~\ref{figMass-vs-m0}. As expected, the mass approaches
the value of $m_{\rm dyn}$ in the chiral limit ($m_0 \to 0$). 
For the model parameters 
used in this paper it reads:
\begin{equation}
m_{\rm dyn} \approx 7.1 \times 10^{-4} \, \Lambda.
\end{equation}

In the same figure, we also
plotted the results for several nonzero values of temperature. These results show that
the value of the dynamical mass in the chiral limit gradually vanishes with increasing
the temperature. Within our numerical accuracy, the corresponding value of the critical
temperature is consistent with the Bardeen-Cooper-Schrieffer theory relation, $T_c \approx 0.57 m_{\rm dyn}$.
We also note that the results for $\mu$ and $\Delta$ are trivial at all temperatures
when $\mu_0=0$.

\subsection{Numerical solutions at $T=0$}
\label{gapT=0numerical}

\begin{figure*}[ht]
\includegraphics[width=.45\textwidth]{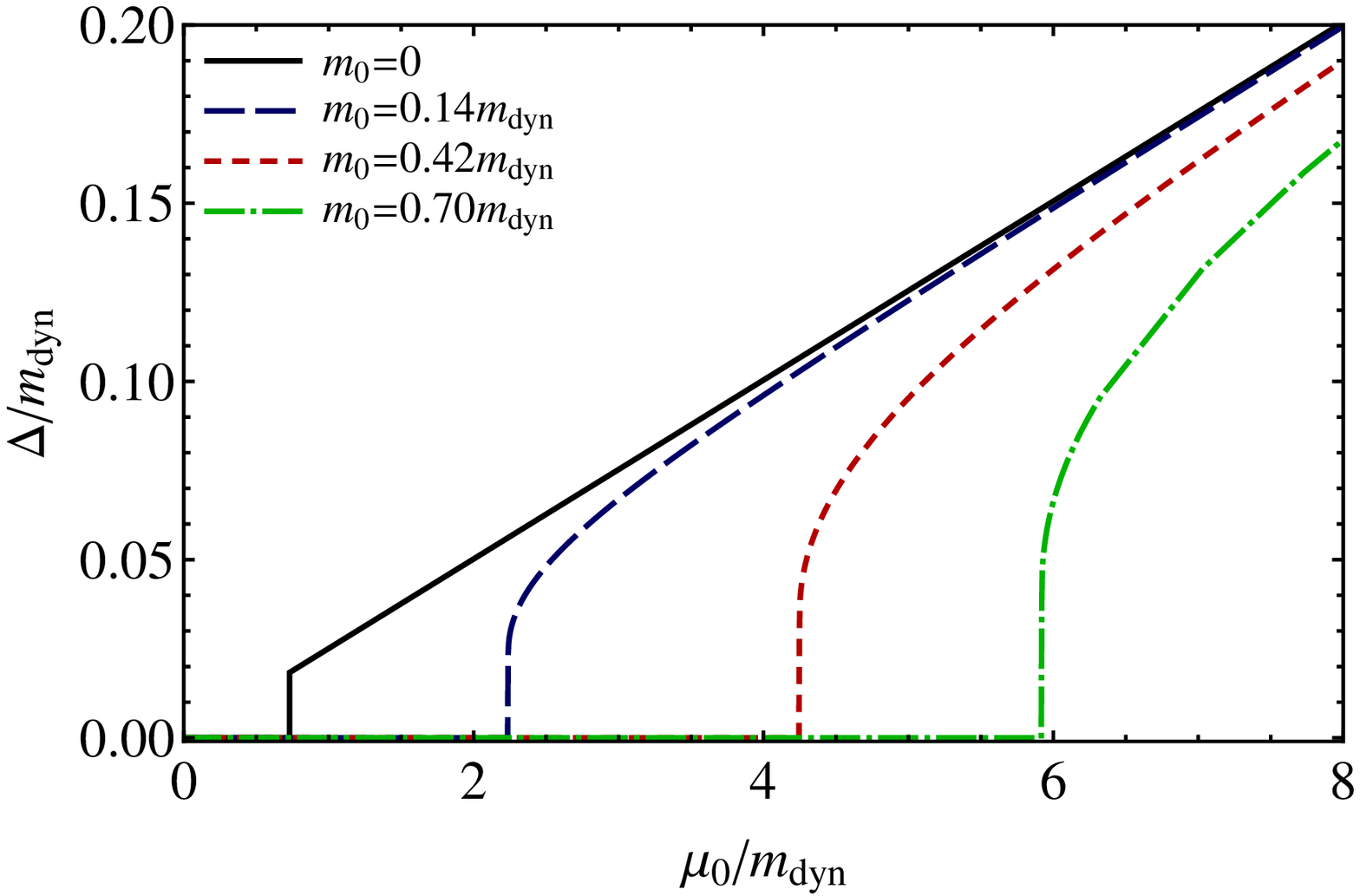}
\hspace{10pt}
\includegraphics[width=.45\textwidth]{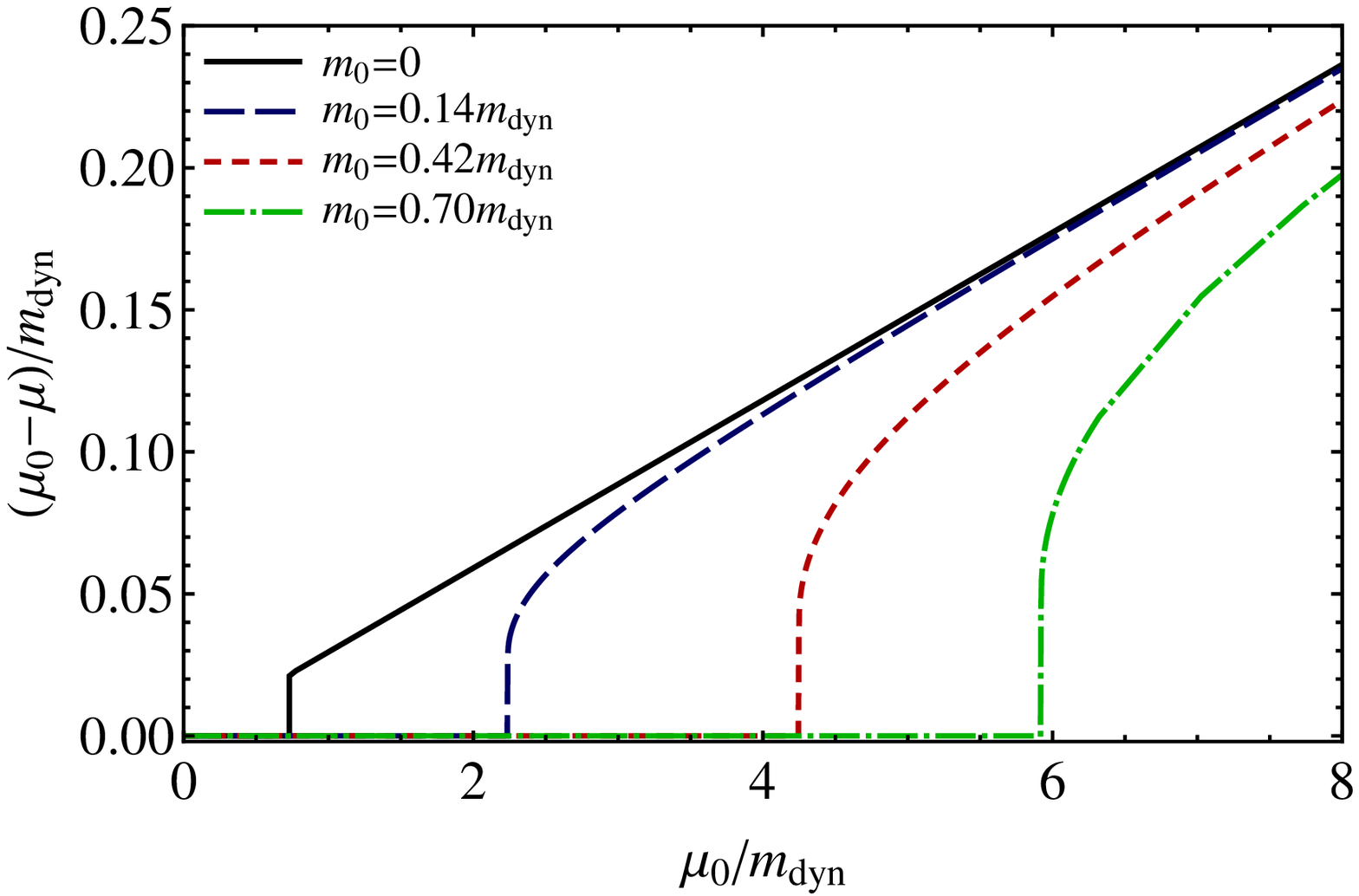}\\
\hspace*{10pt}
\includegraphics[width=.425\textwidth]{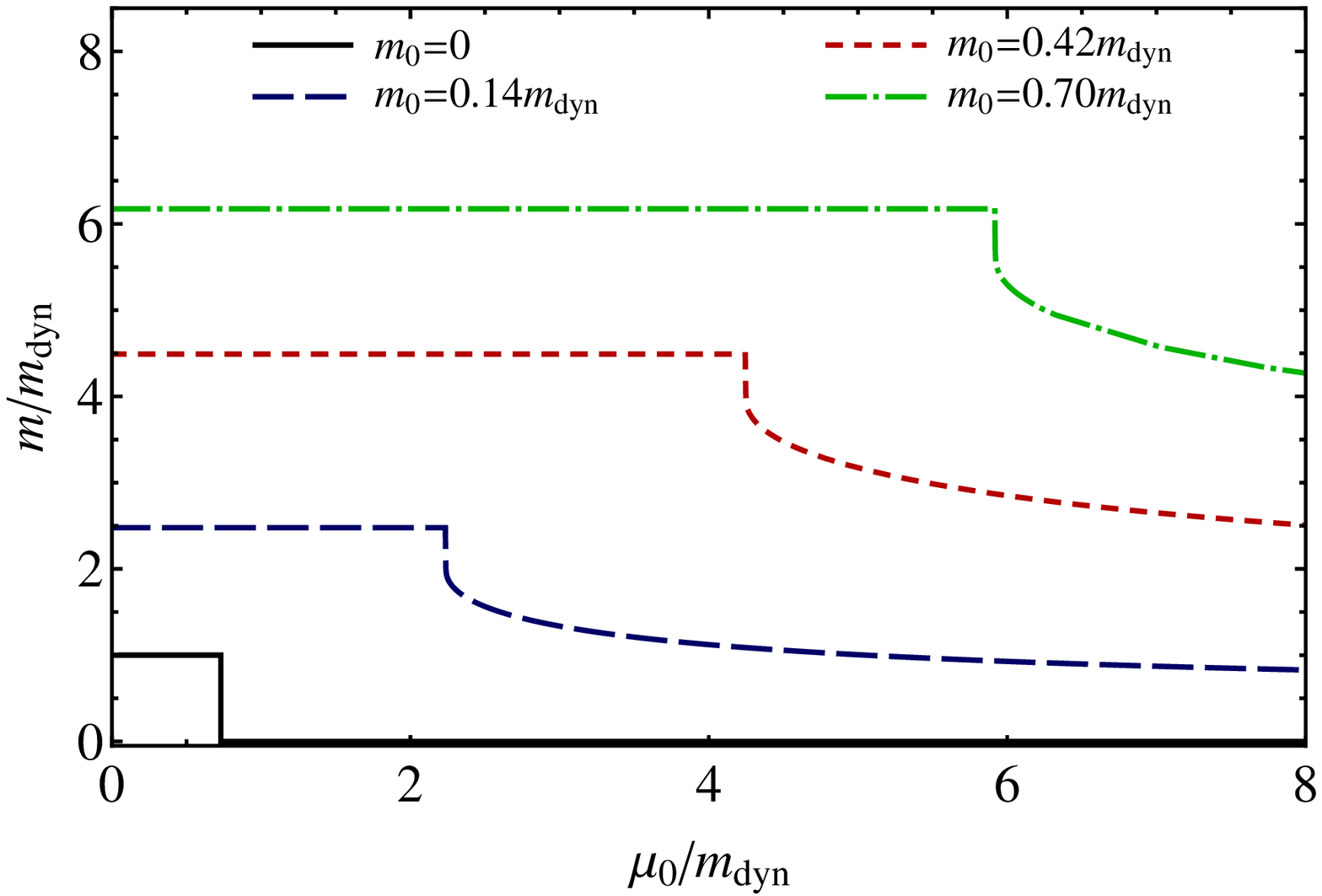}
\hspace{10pt}
\includegraphics[width=.45\textwidth]{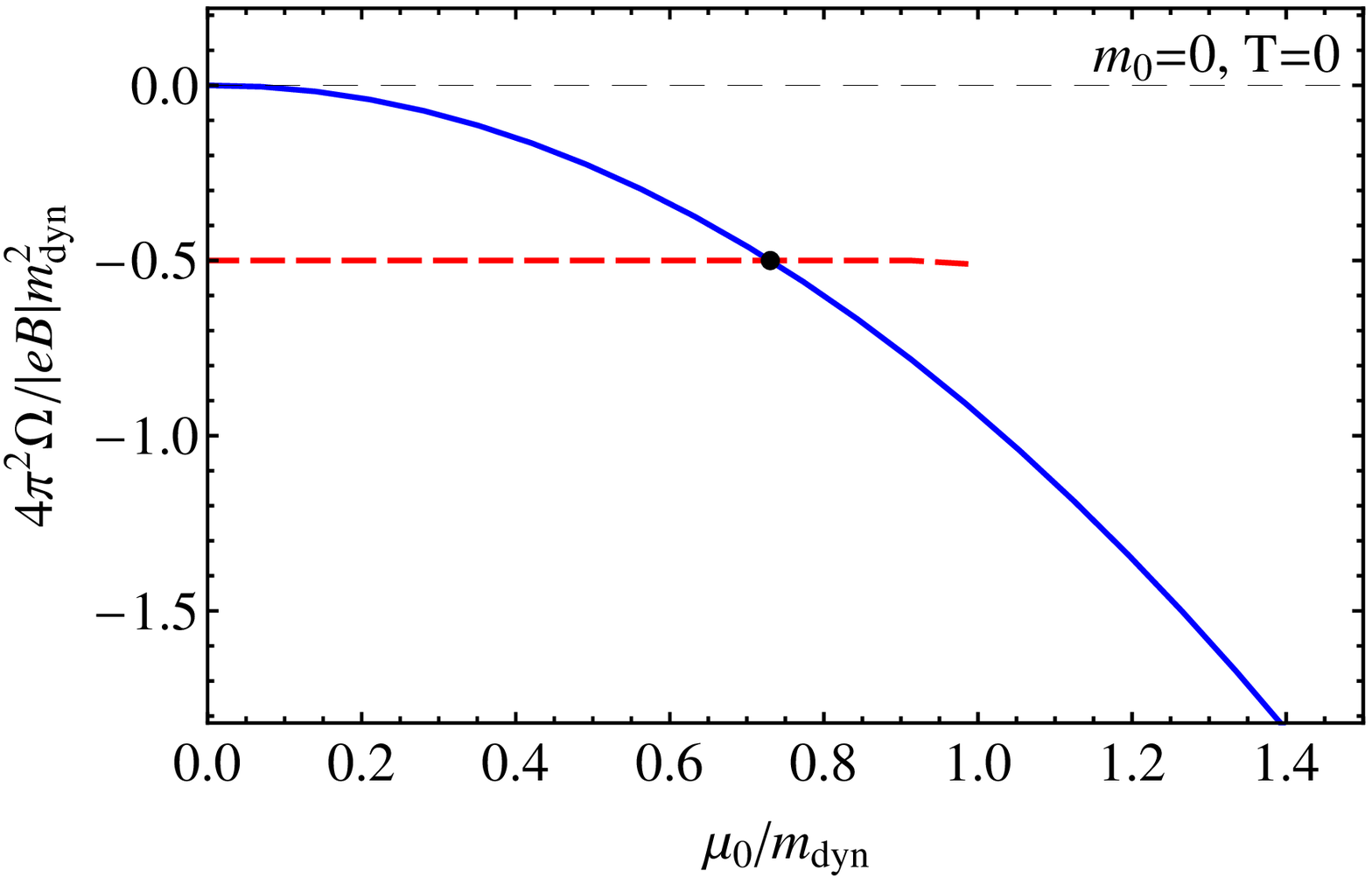}
\caption{(Color online) The zero temperature results for the chiral shift
parameter $\Delta$ (upper left panel), the mass $m$ (lower left panel), and the
difference $\mu_0-\mu$ (upper right panel) on the chemical potential $\mu_0$
for several fixed values of the bare mass: $m_0=0$ (black solid line), 
$m_0=10^{-4}\Lambda=0.14m_{\rm dyn}$ (blue long-dashed line), 
$m_0=3\times 10^{-4}\Lambda=0.42m_{\rm dyn}$ (red short-dashed line), 
$m_0=5\times 10^{-4}\Lambda=0.70m_{\rm dyn}$ (green dash-dotted line). 
The lower right panel shows the free energies for the solution with a nonzero 
dynamical mass (red dashed line) and the solution with a chiral shift (blue solid 
line) in the chiral limit, $m_0=0$, at zero temperature.} 
\label{fig-T0multi-m0}
\end{figure*}

The solutions of $\Delta$ vs $\mu_0$ and $m$ vs $\mu_0$, obtained by the iterating the set
of gap equations, are shown in upper and lower left panels of Fig.~\ref{fig-T0multi-m0},
respectively. There the solutions for several different values of bare masses $m_0$ are shown.
It might be appropriate to note here that, in the vicinity of the phase transitions, we
had obtained a pair of solutions for each fixed value of $m_0$. The two different
solutions are obtained by sweeping over the same range of the chemical potentials $\mu_0$
in two different directions: (i) from left to right and (ii) from right to left. When such
a pair of solution, forming a small hysteresis loop, is observed, a first order phase
transition is expected somewhere within the loop. To determine the location of such a phase
transition, the comparison of free energy densities for the corresponding pairs of solutions
is required. The general expression for the free energy is derived in Appendix~\ref{free-energy}.
By calculating the corresponding expression for each of the solutions around the hysteresis loop,
we could point the position of the actual phase transition. An example of such a calculation
is presented in the lower right panel of Fig.~\ref{fig-T0multi-m0}. There the free energies
of the pair of solutions in the case $m_0=0$ are shown. The blue solid line correspond to the
solution with $m=0$ and
$\Delta\neq 0$. The free energy shown by the red dashed line represents the other solution, in
which the Dirac mass $m$ is nonzero and $\Delta$ is zero at small $\mu_0< m_{\rm dyn}$. The free
energies of the two solutions become equal at $\mu_{0,{\rm cr}}\approx 0.73 m_{\rm dyn}$. This
is where the first order phase transition occurs. Note that the numerical value of
$\mu_{0,{\rm cr}}$ is within several percent of the analytical estimate $m_{\rm dyn}/\sqrt{2}$.

Concerning the solution for $\mu$ vs $\mu_0$, the results are always such that $\mu\approx \mu_0$ to
within a few percent. Therefore, the corresponding plot would give little information. In order to
get a deeper insight into the deviation of $\mu$ from $\mu_0$, we find it instructive to plot the
result for the difference $\mu_0-\mu$ instead. Note that as follows from Eqs.~(\ref{gap-mu-text}) 
and (\ref{A-density-text}), the latter is proportional to the fermion charge density. The result is 
presented in the upper right panel of Fig.~\ref{fig-T0multi-m0}. We see that $\mu_0-\mu$ is always 
positive, meaning that the value of $\mu$ is slightly smaller than $\mu_0$. 

By comparing that graph for $\mu_0-\mu$ with the dependence of the chiral shift
parameter $\Delta$ on $\mu_0$ in the upper left panel of the same figure, we observe that they
have the same qualitative behaviors. 
In particular, $\Delta$ is nonzero in the ground state only if the fermion charge 
density is also nonzero there. In other words, the shift parameter is a manifestation of dynamics in
a system with matter. Note that when the numerical values of the model parameters are used, the
results for $\mu$ and $\Delta$ become $\mu\simeq \mu_0-X\mu_0$ and $\Delta \simeq Y\mu_0$ 
for $m_0 = 0$, where $X\approx 0.0295 $ and $Y\approx 0.0252$.

\subsection{Numerical solutions at $T\neq 0$ and $\mu_0\neq 0$}
\label{gapTnumerical}

\begin{figure*}[ht]
\includegraphics[width=.45\textwidth]{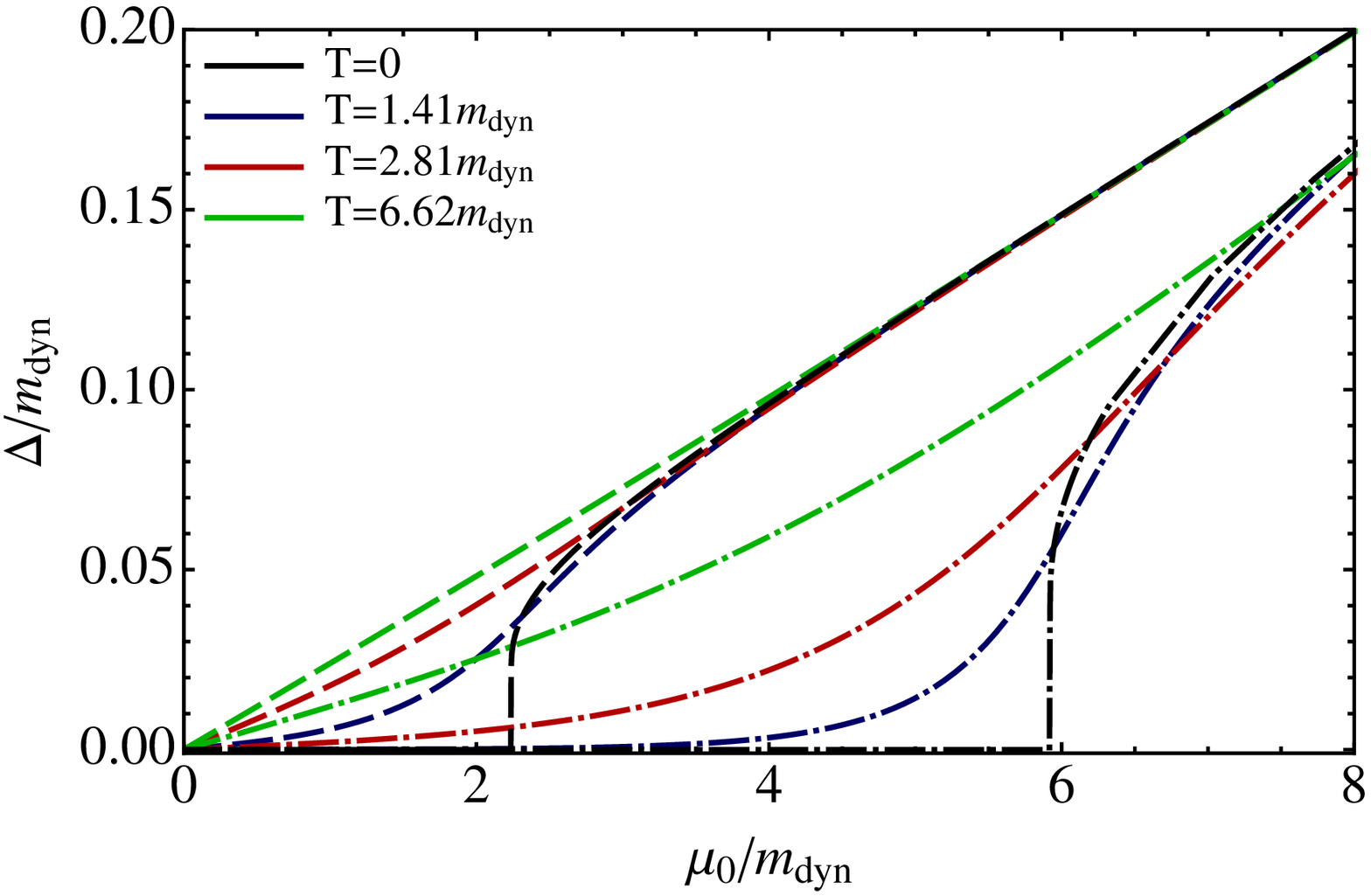}
\hspace*{10pt}
\includegraphics[width=.425\textwidth]{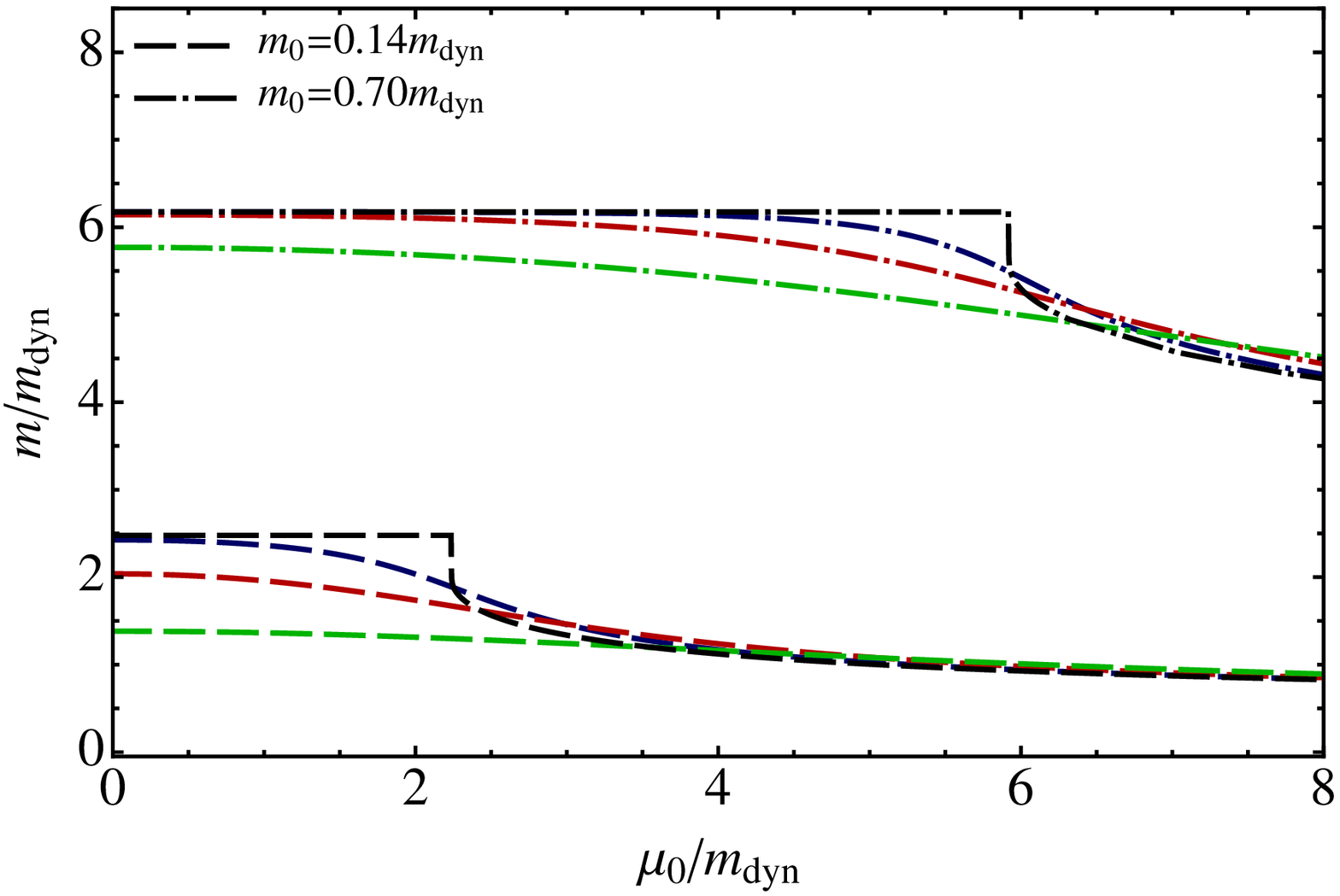}
\caption{(Color online) The nonzero temperature results for the chiral shift parameter
$\Delta$ (left panel) and the mass $m$ (right panel) as function of the chemical potential
$\mu_0$ for several fixed values of the temperature,
$T=0$ (black),
$T=1.41m_{\rm dyn}$ (blue lines),
$T=2.81m_{\rm dyn}$ (red lines), and
$T=5.62m_{\rm dyn}$ (green lines), and two values of the bare mass,
$m_0=0.14m_{\rm dyn}$ (solid lines) and
$m_0=0.70m_{\rm dyn}$ (short-dashed lines).}
\label{fig-Tmulti-mo}
\end{figure*}

Let us now proceed with the numerical solution of the gap equation at nonzero temperature.
At vanishing value of $\mu_0$, several results for the constituent mass have already been
presented in Fig.~\ref{figMass-vs-m0}. The other two parameters, $\mu$ and $\Delta$, were
identically zero in that special case. Here we extend the solutions to nonzero values of
$\mu_0$. The numerical results for $\Delta$ vs $\mu_0$ and $m$ vs $\mu_0$ are presented
in Fig.~\ref{fig-Tmulti-mo}. Note that the dependence $\mu - \mu_0$ vs $\mu_0$ (not shown 
in that figure) is similar to that of $\Delta$ vs $\mu_0$ at all temperatures. As should be expected,
temperature suppresses the dynamical fermion mass (see the right panel of this figure). However,
the situation is quite different for the chiral shift parameter. As one can see in the left
panel of the figure, $\Delta$ is rather insensitive to temperature when $T \ll \mu_0$, and 
{\it increases} with $T$ when $T > \mu_0$. This property reflects the fact that higher 
temperature leads to higher matter density, which is apparently a more favorable environment 
for generating the chiral shift $\Delta$. While the first regime with $T \ll \mu_0$ is appropriate 
for stellar matter, the second one with $T > \mu_0$ (actually, $T \gg \mu_0$) is realized in 
heavy ion collisions. As we discuss in Sec.~\ref{Summary} below, the generation 
of $\Delta$ may have important implications for both stellar matter and 
heavy ion collisions.

\section{Axial current density}
\label{sec-axial-current}

It is instructive to calculate the ground state expectation value of the axial current density.
As we see from Eq.~(\ref{D-axial-current-text}),  it coincides with the function ${\cal D}$,
\begin{eqnarray}
\langle j_5^3\rangle &=& -\tr\left[\gamma^3\gamma^5 G(u,u)\right]={\cal D} .
\label{axial-general}
\end{eqnarray}
In the case of the vanishing Dirac mass, $m=0$, an explicit expression for ${\cal D}$ within 
the momentum cutoff and the proper-time regularization schemes were presented in Eqs.~(\ref{DT0m0}) 
and (\ref{DT0m0-proper-time}), respectively. Both expressions can be written in the same form:
\begin{equation}
\langle j_5^3\rangle \simeq \frac{-eB}{2\pi^2}\left[\mu -2 a s_{\perp}\Delta (\Lambda l)^2 \right],
\label{axial}
\end{equation}
where $a$ is a dimensionless constant of order $1$, determined by the specific regularization scheme. 
When the proper time is used, we find from Eq.~(\ref{DT0m0-proper-time}) that $a=1/4$. In the case 
of the cutoff regularization, it is defined by Eq.~(\ref{sumkappa}). Note that qualitatively the same 
result is also obtained in the point-splitting regularization \cite{Gorbar:2010}.

The first term in the parenthesis in Eq.~(\ref{axial}) is the same topological term that was derived in the free 
theory in Ref.~\cite{Metlitski:2005pr}, while the second term is an outcome of interactions \cite{Gorbar:2009bm}.
It is interesting to note that, by making use of the gap equation (\ref{gap-Delta-text}) for $\Delta$,
the result for the axial current can be also rewritten in an alternative form:
\begin{equation}
\langle j_5^3\rangle =-\frac{2\Delta}{G_{\rm int} } =- \frac{\Delta}{2\pi^2}\frac{\Lambda^2}{g}.
\label{axial-Delta}
\end{equation}
While this may not be very convenient in the free theory, in which 
both the coupling constant $g$ and the chiral shift $\Delta$ vanish,
and the cutoff is formally infinite, it is helpful to get a deeper insight in interacting theory.

Formally, the results for the the axial current either in Eq.~(\ref{axial}) or in Eq.~(\ref{axial-Delta}) appear
to be quadratically divergent when $\Lambda\to \infty$. It should be noticed, however, that the solution to 
the gap equation, see Eq.~(\ref{Delta-cutoff}) in the case of cutoff regularization and  Eq.~(\ref{Delta-proper-time}) 
in the case of proper time regularization, is inversely proportional to $\Lambda^2$, i.e., 
$\Delta\sim g \mu eB/\Lambda^2$. Taking this into account, we see that the axial current density is 
actually finite in the continuum limit $\Lambda\to \infty$,
\begin{equation}
\langle j^{3}_5 \rangle \simeq \frac{-eB}{2\pi^{2}} \mu + a \frac{\Lambda^2}{\pi^{2}}  \Delta
\simeq \frac{-eB}{2\pi^{2}}\frac{\mu}{\left(1+ 2 a g\right)} .
\label{a}
\end{equation}

Before concluding this section, let us also note the following expression for fermion number density:
\begin{eqnarray}
\langle j^0\rangle &=& -\tr\left[\gamma^0 G(u,u)\right] = {\cal A} .
\end{eqnarray}
The explicit form of the function ${\cal A}$ is derived in Appendix~\ref{AppPropagator}. The corresponding
result is complicated and adds no new information when the solution to the gap equation is available.
Indeed, the fermion number density can be conveniently rewritten in a simpler form by making use of
the gap equation (\ref{gap-mu-text}) for $\mu$,
\begin{equation}
\langle j^0\rangle = 2\frac{\mu_0-\mu}{G_{\rm int} } = \frac{\mu_0-\mu}{2\pi^2}\frac{\Lambda^2}{g} .
\end{equation}
This shows that the result for this density is proportional to $\mu_0-\mu$ presented earlier.

In the case of a strong magnetic field, when the LLL approximation is appropriate, we find that the 
chiral shift parameter (and, thus, the axial current density) and the fermion number density are 
proportional to each other. This is apparent in Fig.~\ref{fig-T0multi-m0}. The underlying reason 
for this proportionality is the same (up to a sign)  LLL contribution to both functions ${\cal A}$ 
and ${\cal D}$. Moreover, this property seems to be at least approximately
valid in a general case. In turn, this suggests
that a fermion number density and the chiral shift parameter are two closely connected
characteristics of the normal phase of magnetized relativistic matter.

\section{Discussion and summary}
\label{Summary}

\subsection{Fermi surface}

\begin{figure}
\begin{center}
\includegraphics[width=.4\textwidth]{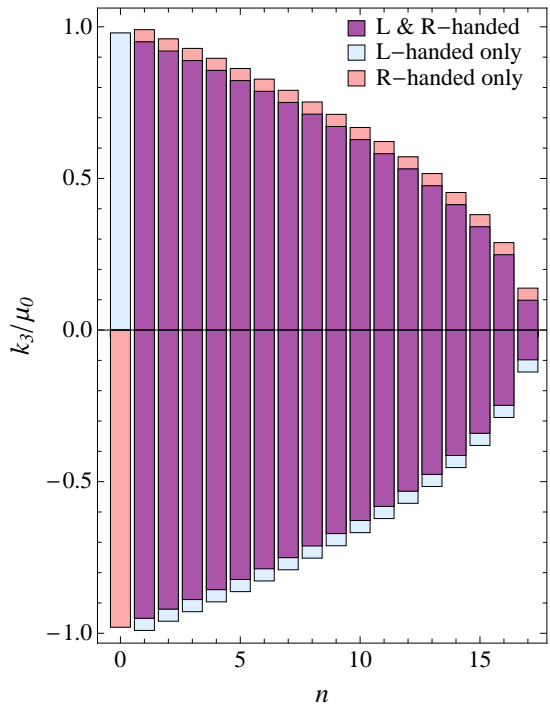}
\caption{(Color online) A schematic distribution of (negatively charged) particles in the ground state of 
dense relativistic matter in a magnetic field (pointing in the positive $z$ direction). The values of the 
quantum number $n$ (Landau levels) are shown along the horizontal axis, while the longitudinal momenta are shown along the vertical axis.
The colored bars indicate the filled states of given chirality.}
\label{figFermiSurface}
\end{center}
\end{figure}

The immediate implication of a nonzero chiral shift parameter in dense magnetized matter is the 
modification of the quasiparticle dispersion relations, see Eqs.~(\ref{omega0}) and (\ref{omegan}).  
These relations can be used to define the ``Fermi surface" in the space of the longitudinal momentum 
$k^3$ and the Landau index $n$. Note that the quantity $2n |eB|$ plays the role analogous 
to the square of the transverse momentum $k_\perp^2\equiv (k^1)^2+(k^2)^2$ in the absence of 
the magnetic field. Following the standard philosophy, we define the Fermi surface as the hypersurface 
in the space of quantum numbers $n$ and $k^3$, which correspond to quasiparticles with zero energy, i.e.,
\begin{eqnarray}
n=0: &\!\!\!\!& k^{3}=\pm\sqrt{(\mu-s_\perp \Delta)^2-m^2},   \label{Fermi-k3-1}\\
n>0: &\!\!\!\!& k^{3}=\pm\sqrt{\left(\sqrt{\mu^2-2n|eB|}\pm s_\perp \Delta\right)^2-m^2}. \label{Fermi-k3-2}
\end{eqnarray}
(In the last equation, all four combinations of signs are possible.) In order to better 
understand the nature of quasiparticles at the Fermi surface, described by Eqs.~(\ref{Fermi-k3-1}) 
and (\ref{Fermi-k3-2}), we recall that there are two types of quasiparticles. The Dirac structures of 
their wave functions are obtained by applying the projection operators in Eq.~(\ref{projectorH}). 
In relativistic dense matter ($\mu\gg m$), the corresponding states at the Fermi surface can be 
approximately characterized by their chiralities. This follows from the fact that $|k^3|\gg m$ for 
a large fraction of the Fermi surface in Eq.~(\ref{Fermi-k3-2}), except for the limiting values of $n$ 
around $n_{\rm max} \equiv \left[\mu^2/(2|eB|)\right]$. At such large values of the relative momentum, 
the projection operators in Eq.~(\ref{projectorH}) are very closely related to the chiral projectors. 
Indeed, for $|k^3|\gg m$, the relation between the two sets of projectors is approximately the same 
as in the {\em massless} case in Eq.~(\ref{Hpm-P5pm}). Taking this into account, it is possible to 
define quasiparticles at the Fermi surface, which are predominantly left-handed or right-handed.
Without loss of generality, let us assume that $\sign(eB)>0$. Then, the Fermi surface for the 
{\em predominantly left-handed} particles is given by 
\begin{eqnarray}
n=0: &\!\!\!\!& k^{3}=+\sqrt{(\mu-s_\perp \Delta)^2-m^2},   \label{Fermi-k3-1L}\\
n>0: &\!\!\!\!& k^{3}=+\sqrt{\left(\sqrt{\mu^2-2n|eB|}- s_\perp \Delta\right)^2-m^2}, \label{Fermi-k3-2L} \\
        &\!\!\!\!& k^{3}=-\sqrt{\left(\sqrt{\mu^2-2n|eB|}+ s_\perp \Delta\right)^2-m^2}, \label{Fermi-k3-3L}
\end{eqnarray}
and the Fermi surface for the {\em predominantly right-handed} particles is 
\begin{eqnarray}
n=0: &\!\!\!\!& k^{3}=-\sqrt{(\mu-s_\perp \Delta)^2-m^2},   \label{Fermi-k3-1R}\\
n>0: &\!\!\!\!& k^{3}=-\sqrt{\left(\sqrt{\mu^2-2n|eB|}- s_\perp \Delta\right)^2-m^2}, \label{Fermi-k3-2R}\\
        &\!\!\!\!& k^{3}=+\sqrt{\left(\sqrt{\mu^2-2n|eB|}+ s_\perp \Delta\right)^2-m^2}. \label{Fermi-k3-3R}
\end{eqnarray}
In the massless case, of course, this correspondence becomes exact. Then, we find that the Fermi surface 
for fermions of a given chirality is asymmetric in the direction of the magnetic field. In Fig.~\ref{figFermiSurface}, 
we show a schematic distribution of negatively charged fermions and take into account that the parameter 
$s_\perp \Delta$ has the same sign as the chemical potential, see Eqs.~(\ref{Delta-cutoff}) or (\ref{Delta-proper-time}). 
(A similar distribution is also valid for positively charged fermions, but the left-handed and right-handed fermions 
will interchange their roles.) For the fermions of a given chirality, the LLL and the higher Landau levels give 
opposite contributions to the overall asymmetry of the Fermi surface. For example, the left-handed electrons in 
the LLL occupy only the states with {\em positive} longitudinal momenta (pointing in the magnetic field direction). The 
spins of the corresponding LLL electrons point against the magnetic field direction. In the higher Landau levels, 
while the left-handed electrons can have both positive and negative longitudinal momenta (as well as both spin 
projections), there are more states with {\em negative} momenta occupied, see Fig.~\ref{figFermiSurface}. 
If there are many Landau levels occupied, which is the case when $\mu\gg \sqrt{|eB|}$, the relative contribution 
of the LLL to the whole Fermi surface is small, and the overall asymmetry is dominated by higher Landau levels.
In the opposite regime of superstrong magnetic field (if it can be realized in compact stars at all), only the LLL is 
occupied and, therefore, the overall asymmetry of the Fermi surface will be reversed. In the intermediate regime 
of a few Landau levels occupied, one should expect a crossover from one regime to the other, where the 
asymmetry goes through zero.

\subsection{Effects in compact stars}

The asymmetry with respect to longitudinal momentum $k^{3}$ of the opposite chirality fermions in the 
ground state of dense magnetized matter, discussed in the preceding subsection, may have 
important physical consequences. For example, the fact that only the left-handed fermions participate 
in the weak interactions means that the neutrinos will scatter asymmetrically off the matter, in 
which the chiral shift parameter is nonvanishing. 

By making use of this observation, a qualitatively new mechanism for the pulsar kicks \cite{pulsarkicks} 
was proposed in Ref.~\cite{Gorbar:2009bm}. It can be realized in almost any type of relativistic matter 
inside a protoneutron star (e.g., the electron plasma of the nuclear/hadronic matter, or the quark and 
electron plasmas in the deconfined quark matter), in which a nonzero chiral shift parameter $\Delta$ 
develops. 

When the original trapped neutrinos
gradually diffuse through the bulk of a protoneutron star, they can {\em build up} an asymmetric momentum 
distribution as a result of their multiple elastic scattering on the {\em nonisotropic} state of left-handed fermions 
(electrons or quarks). This is in contrast to the common dynamics of diffusion through an {\em isotropic} hot matter, 
which leads to a very efficient thermal isotropization and, therefore, a wash out of any original nonisotropic 
distribution of neutrinos \cite{Kusenko,SagertSchaffner}.

It appears also very helpful for the 
new pulsar kick mechanism that the chiral shift parameter is not much affected even by moderately 
high temperatures, $10~\mbox{MeV} \lesssim T\lesssim 50~\mbox{MeV}$, present during the earliest
stages of protoneutron stars \cite{AnnRevNuclPart}. Indeed, as our findings show, the value of 
$\Delta$ is primarily determined by the chemical potential and has a weak/nonessential temperature 
dependence when $\mu\gg T$. In the stellar context, this ensures the feasibility of the proposed mechanism 
even at the earliest stages of the protoneutron stars, when there is sufficient amount of thermal energy 
to power the strongest (with $v\gtrsim 1000~\mbox{km/s}$) pulsar kicks observed \cite{pulsarkicks}.
Alternatively, the constraints of the energy conservation would make it hard, if not impossible, to explain 
any sizable pulsar kicks if the interior matter is cold ($T\lesssim 1~\mbox{MeV}$).
 
Let us also mention that the robustness of the chiral shift in hot magnetized matter may be useful to 
provide an additional neutrino push to facilitate successful supernova explosions as suggested in 
Ref.~\cite{Fryer:2005sz}. The specific details of such a scenario are yet to be worked out.

\subsection{Heavy ion physics}

It is natural to ask whether the chiral shift parameter can have any interesting implications in the regime 
of relativistic heavy ion collisions. As was recently discussed in the literature, hot relativistic matter 
in a magnetic field may have interesting properties even in the absence of the chiral shift parameter. 
The examples of the recently suggested phenomena, that appear to be closely related to the generation 
of the chiral shift, are the chiral magnetic effect \cite{Kharzeev:2007tn,Rebhan,Fukushima}, the chiral 
magnetic spiral \cite{Basar:2010zd,Kim,Frolov}, and the chiral magnetic wave \cite{Kharzeev}.

 As we find in this study, at high temperatures, i.e., in the regime relevant for relativistic heavy ion 
collisions, the chiral shift parameter is generated for any nonzero chemical potential. This is seen 
from the results presented in Fig.~\ref{fig-Tmulti-mo}. However, its role is not as obvious as in the case of
stellar matter. At high temperatures, the Fermi surface and the low-energy excitations in its vicinity 
are not very useful concepts any more. Instead, it is the axial current itself that is of interest.
The chiral shift should induce a correction to the topological axial current (\ref{MZ}). As seen 
from Eq.~(\ref{axial}), the corresponding correction in the NJL model studied here is proportional 
to the chiral shift parameter $\Delta$, multiplied by a factor $(\Lambda l)^2$, where $\Lambda$ is 
the ultraviolet cutoff. Formally, the product of $\Delta$ and $(\Lambda l)^2$ is finite and is 
proportional to the chemical potential. However, unlike the topological term, which is also 
proportional to the chemical potential, the dynamical one contains an extra factor of the 
coupling constant. Therefore, only at relatively strong coupling, which can be provided by 
QCD interactions, the effect of the chiral shift parameter on the axial current can be substantial.
 
Following the ideas similar to those that were used in the chiral magnetic effect \cite{Kharzeev:2007tn,Rebhan,Fukushima}, 
we would like to suggest that the axial current by itself can play an important role in hot matter produced by 
heavy ion collisions. It can lead to a modified version of the chiral magnetic effect, which does not rely 
on the initial topological charge fluctuations. This can presumably be realized as follows. An initial axial 
current generates an excess of opposite chiral charges around the polar regions of the fireball. Then, 
these chiral charges trigger two ``usual" chiral magnetic effects with opposite directions of the vector 
currents at the opposite poles. The inward flows of these electric currents will diffuse inside the fireball, 
while the outward flows will lead to a distinct observational signal: an excess of same sign charges going 
back-to-back. 

Concerning the regime of hot relativistic matter, let us also mention that it will be of interest to extend 
our analysis of magnetized relativistic matter to address the properties of collective modes, similar to 
those presented in Ref.~\cite{Kharzeev}, by studying various current-current correlators.

\subsection{Renormalizability vs nonrenormalizability}

The present analysis was realized in the NJL model. It would be important to extend
it to renormalizable field theories, especially, QED and QCD. In connection with
that, we would like to note the following. The expression for the chiral shift parameter,
$\Delta \sim g\mu \, eB/\Lambda^2$, obtained in the NJL model implies that both fermion
density and magnetic field are necessary for the generation of $\Delta$. This feature
should also be valid in renormalizable theories. As for the cutoff $\Lambda$, it enters
the results only because of the nonrenormalizability of the NJL model.

Similar studies of chiral
symmetry breaking in the vacuum ($\mu_0 = 0$) QED and QCD in a magnetic field show
that the cutoff scale $\Lambda$ is replaced by $\sqrt{|eB|}$ there \cite{QED}. Therefore,
one might expect that in QED and QCD with both $\mu$ and $B$ being nonzero, $\Lambda$
will be replaced by a physical parameter, such as $\sqrt{|eB|}$. This in turn suggests that
a constant chiral shift parameter $\Delta$ will become a running quantity that depends on the
longitudinal momentum $k^3$ and the Landau level index $n$.

Another important feature that one could expect in QCD in a magnetic field is a topological
contribution in the baryon charge \cite{SS} connected with collective massless fermion
excitations in the phase with spontaneous chiral symmetry breaking. This feature could
dramatically change the properties of that phase \cite{Preis}.

It is clear that dynamics in dense relativistic matter in a magnetic field is rich and
sophisticated. In particular, one could expect surprises in studies of the phase diagram
of QCD in a magnetic field \cite{Mizher,D'Elia,Gatto:2010qs,Fayazbakhsh:2010bh}.

\begin{acknowledgments}
The authors would like to thank V.~P.~Gusynin for fruitful discussions and A.~Schmitt 
for useful comments. The work of E.V.G. was supported partially by the SCOPES under 
Grant No. IZ73Z0-128026 of the Swiss NSF, under Grant No. SIMTECH 246937 
of the European FP7 program, the joint Grant RFFR-DFFD No. F28.2/083 of the 
Russian Foundation for Fundamental Research and of the Ukrainian State Foundation 
for Fundamental Research (DFFD). The work of V.A.M. was supported by the Natural 
Sciences and Engineering Research Council of Canada. The work of I.A.S. is supported 
in part by a start-up fund from the Arizona State University and by the U.S. National 
Science Foundation under Grant No. PHY-0969844.
\end{acknowledgments}

\appendix

\begin{widetext}

\section{The full fermion propagator}
\label{AppPropagator}

\subsection{General result}

In this appendix we calculate the explicit form of the full fermion propagator. {From} the 
definition of the inverse propagator in Eq.~(\ref{ginverse}), it follows that 
\cite{footnote-charge}
\begin{eqnarray}
G(u,u^{\prime}) &=& i\langle u|\left[(i\partial_t+\mu)\gamma^0 -
(\bm{\pi}_{\perp}\cdot\bm{\gamma})-\pi^{3}\gamma^3
+i\tilde{\mu}\gamma^1\gamma^2+i\Delta\gamma^0\gamma^1\gamma^2
-m\right]^{-1}|u^\prime\rangle
\nonumber\\
&=& i\langle u| \left[(i\partial_t+\mu)\gamma^0 -
(\bm{\pi}_{\perp}\cdot\bm{\gamma})-\pi^{3}\gamma^3 +i\tilde{\mu}\gamma^1\gamma^2
-i\Delta\gamma^0\gamma^1\gamma^2 +m\right]
\nonumber\\
&\times & \Big\{\left[ (i\partial_t+\mu)\gamma^0 -
(\bm{\pi}_{\perp}\cdot\bm{\gamma})-\pi^{3}\gamma^3 +i\tilde{\mu}\gamma^1\gamma^2+
i\Delta\gamma^0\gamma^1\gamma^2-m\right]
\nonumber\\
&& \times \left[ (i\partial_t+\mu)\gamma^0 -
(\bm{\pi}_{\perp}\cdot\bm{\gamma})-\pi^{3}\gamma^3 +i\tilde{\mu}\gamma^1\gamma^2
-i\Delta\gamma^0\gamma^1\gamma^2+
m\right]\Big\}^{-1}|u^\prime\rangle
\nonumber\\
&=& i\langle u| \left[(i\partial_t+\mu)\gamma^0 -
(\bm{\pi}_{\perp}\cdot\bm{\gamma})-\pi^{3}\gamma^3 +i\tilde{\mu}\gamma^1\gamma^2
-i\Delta\gamma^0\gamma^1\gamma^2+m\right]
\nonumber\\
&\times& \left[(i\partial_t+\mu+i\tilde{\mu}\gamma^0\gamma^1\gamma^2)^2- \bm{\pi}_{\perp}^2 -(\pi^{3})^2
-(m-i\Delta\gamma^0\gamma^1\gamma^2)^2
-ie B \gamma^1\gamma^2
-2i\gamma^1\gamma^2(\tilde{\mu}+\Delta\gamma^0)\pi^{3}\gamma^3\right]^{-1}
|u^\prime\rangle,\nonumber\\
\label{propagator1}
\end{eqnarray}
where $u=(t,\mathbf{r})$ and $\mathbf{r}=(x,y,z)$. Note that the canonical momenta 
are $\pi_{\perp}^{k} \equiv i \partial^k + e A^k$ (with $k=1,2$) and $\pi^{3} =- i \partial_3$, and the 
Dirac structure $\gamma^3\gamma^5$ is rewritten in an equivalent form, 
$\gamma^3\gamma^5=i\gamma^0\gamma^1\gamma^2$.

In order to derive an expression for the propagator in a form of an expansion over the Landau levels,
we follow the same approach as in Ref.~\cite{GGMS2008}. We start by switching to the Fourier
transform in time $t$ and in $z$-coordinate (i.e., the coordinate along the magnetic field),
\begin{eqnarray}
G(\omega,k^{3};\mathbf{r},\mathbf{r}^{\prime})
=\int dt \, dz\, e^{i\omega (t-t^\prime)-ik^{3} (z-z^\prime)}G(u,u^\prime)
=i\left[W-(\bm{\pi}_{\perp,\mathbf{r}}\cdot{\pmb\gamma})\right]
\langle\mathbf{r} | \left({\cal M}-\bm{\pi}_{\perp}^{2}- ie B \gamma^1\gamma^2\right)^{-1} |
\mathbf{r}^{\prime}\rangle,
\label{propagator-FT}
\end{eqnarray}
where $\bm{\pi}_{\perp,\mathbf{r}}$ is the differential operator of the canonical momentum in the
coordinate space spanned by vector $\mathbf{r}$. The explicit structure of $W$ and ${\cal M}$,
which are matrices in Dirac space, follows directly from Eq.~(\ref{propagator1}), i.e.,
\begin{eqnarray}
W&=&(\omega+\mu)\gamma^{0}+i\tilde{\mu}\gamma^1\gamma^2
-i\Delta\gamma^0\gamma^1\gamma^2+m-k^{3}\gamma^3,
\label{matrices-MW}\\
{\cal M} &=& (\omega+\mu+i\tilde{\mu}\gamma^0\gamma^1\gamma^2)^2
-(m-i\Delta\gamma^0\gamma^1\gamma^2)^2-(k^{3})^2
-2i\gamma^1\gamma^2(\tilde{\mu}+\Delta\gamma^0)k^{3}\gamma^3.
\label{matrices-WM}
\end{eqnarray}
By noting that all three operators ${\cal M}$, $\bm{\pi}_{\perp}^{2}$ and $ie B \gamma^1\gamma^2$
inside the matrix element on the right hand side of Eq.~(\ref{propagator-FT}) commute,
we proceed to build their common basis of eigenfunctions. We start from the operator
$\bm{\pi}_{\perp}^{2}$. As is well known, it has the eigenvalues $(2n+1)|e B |$ with $n=0,1,2,\dots$.
The corresponding normalized wave functions in the Landau gauge,  $\mathbf{A}=(0, B x)$, read
\begin{eqnarray}
\psi_{np}(\mathbf{r})\equiv \langle \mathbf{r}  | n p \rangle
=\frac{1}{\sqrt{2\pi l}}\frac{1}{\sqrt{2^nn!\sqrt{\pi}}}
H_n\left(\frac{x}{l}+pl\right)e^{-\frac{1}{2l^2}(x+pl^2)^2} e^{i s_{\perp}py},
\end{eqnarray}
where $H_{n}(x)$ are the Hermite polynomials \cite{GR} and $l=1/\sqrt{|e B |}$ is the
magnetic length. These wave functions satisfy the conditions of normalizability and
completeness,
\begin{eqnarray}
\int d^{2}\mathbf{r} \,\langle n p |\mathbf{r} \rangle \langle \mathbf{r}  | n^{\prime} p^{\prime}  \rangle = 
\int d^{2}\mathbf{r}\,\psi^{*}_{np}(\mathbf{r})\psi_{n^{\prime}p^{\prime}}(\mathbf{r}) &=& \delta_{nn^{\prime}}
\delta(p-p^{\prime}),
\\
\sum\limits_{n=0}^{\infty}\int\limits_{-\infty}^{\infty} dp \,
\langle \mathbf{r}  | n p\rangle \langle n p |\mathbf{r}^{\prime}  \rangle 
=\sum\limits_{n=0}^{\infty}\int\limits_{-\infty}^{\infty} dp\,
\psi_{np}(\mathbf{r}) \psi^{*}_{np}(\mathbf{r}^{\prime}) &=& \delta(\mathbf{r}-\mathbf{r}^{\prime}),
\label{completeness}
\end{eqnarray}
respectively. Then, by making use of the spectral expansion of the unit operator (\ref{completeness}),
we can rewrite the matrix element on the right hand side of Eq.~(\ref{propagator-FT}) as follows:
\begin{eqnarray}
\langle\mathbf{r}|\left( {\cal M}-\bm{\pi}_{\perp}^{2} - ieB\gamma^1\gamma^2\right)^{-1}|
\mathbf{r}^{\prime}\rangle 
&=&\sum\limits_{n=0}^{\infty}\int\limits_{-\infty}^{\infty} dp\,
\langle \mathbf{r}  | n p\rangle 
\left[ {\cal M}-(2n+1)|eB|- ieB\gamma^1\gamma^2\right]^{-1} 
\langle n p| \mathbf{r}^{\prime}  \rangle  \nonumber\\
&= & \frac{e^{i\Phi(\mathbf{r},\mathbf{r}^{\prime})}}
{2\pi l^2} e^{-\xi/2}\sum\limits_{n=0}^\infty\frac{L_{n}(\xi)}{{\cal M}-(2n+1)|e B | - ieB\gamma^{1}\gamma^{2}},
\label{auxillary-prop}
\end{eqnarray}
where $\xi \equiv (\mathbf{r}-\mathbf{r}^{\prime})^2/(2l^2)$ and $\Phi(\mathbf{r},\mathbf{r}^{\prime})$
is the Schwinger phase, whose explicit form is given by
\begin{equation}
\Phi(\mathbf{r},\mathbf{r}^{\prime})
= - e\int\limits_{\mathbf{r}^{\prime}}^{\mathbf{r}}dz_{i}A_{i}(z)
= - s_{\perp}\frac{(x+x^\prime)(y-y^\prime)}{2l^2}.
\end{equation}
This phase has a universal form for charged particles in a constant magnetic field. Its
origin is related to the fact that the commutative group of translations is
replaced by the noncommutative group of the so-called {\em magnetic translations} \cite{Zak}.
In the derivation of Eq.~(\ref{auxillary-prop}), we calculated exactly the integral over
the quantum number $p$  by making use of formula $7.377$ from Ref.~\cite{GR},
\begin{equation}
\int\limits_{-\infty}^\infty\,e^{-x^2}H_m(x+y)H_n(x+z)dx
=2^n\pi^{1/2}m!z^{n-m}L_m^{n-m}(-2yz),
\end{equation}
which assumes $m\le n$. By definition, $L^{\alpha}_n$ are the generalized Laguerre
polynomials, and $L_n \equiv L^{0}_n$ \cite{GR}.

By noticing that the matrix $ieB\gamma^1\gamma^2$ reduces down to its eigenvalues
$\pm |e B |$ in the subspaces defined by the spin projection operators,
\begin{equation}
{\cal P}_{\pm} =\frac{1}{2}\left[1\pm i\gamma^{1}\gamma^{2}\sign (eB)\right],
\end{equation}
the $n$th term in the sum in Eq.~(\ref{auxillary-prop}) can be conveniently decomposed
into the sum of two contributions,
\begin{equation}
\frac{L_{n}(\xi)}{{\cal M}-(2n+1)|e B | - ieB\gamma^{1}\gamma^{2}}
=\frac{{\cal P}_{-}L_{n}(\xi)}{{\cal M}-(2n+1)|e B |+|e B |}+ \frac{{\cal P}_{+}L_{n}(\xi)}{{\cal
M}-(2n+1)|e B |-|e B |}.
\label{A9}
\end{equation}
Here the ordering of the matrix factors ${\cal P}_{\pm}$ and $({\cal M}- 2n|e B |)^{-1}$ is of
no importance because ${\cal M}$ commutes with both projectors.
By substituting the last expression into Eq.~(\ref{auxillary-prop}) and redefining the summation
index $n\to n-1$ in the second term, the result for the matrix element can be written in a
compact form as
\begin{eqnarray}
\langle\mathbf{r}|\left[{\cal M}-\bm{\pi}_{\perp}^{2}-ieB\gamma^1\gamma^2\right]^{-1}|
\mathbf{r}^{\prime}\rangle=\frac{e^{i\Phi(\mathbf{r},\mathbf{r}^{\prime})}}{2\pi l^{2}}
e^{-\xi/2}\sum\limits_{n=0}^\infty\frac{{\cal P}_{-}L_{n}(\xi)+{\cal P}_{+}L_{n-1}(\xi)}{{\cal
M}- 2n|e B |},
\end{eqnarray}
where $L_{-1}\equiv 0$  by definition. Finally, by noting that
\begin{eqnarray}
\pi_{x}e^{i\Phi} &=& e^{i\Phi}\left(-i\partial_{x}- s_\perp  \frac{y-y^{\prime}}{2l^{2}}\right),\\
\pi_{y}e^{i\Phi} &=& e^{i\Phi}\left(-i\partial_{y}+ s_\perp \frac{x-x^{\prime}}{2l^{2}}\right),
\end{eqnarray}
the full propagator (\ref{propagator-FT}) takes the form of a product of
the Schwinger phase factor and a translation invariant part, i.e.,
\begin{equation}
G(\omega,k^{3};\mathbf{r},\mathbf{r}^{\prime})=e^{i\Phi(\mathbf{r},\mathbf{r}^{\prime})}
\bar{G}(\omega,k^{3};\mathbf{r}-\mathbf{r}^{\prime}),
\end{equation}
where the translation invariant part of the propagator is
\begin{equation}
\bar{G}(\omega,k^{3};\mathbf{r}-\mathbf{r}^{\prime}) = i\left[W-\gamma^{1}
\left(-i\partial_{x}- s_\perp \frac{y-y^{\prime}}{2l^{2}}\right)-\gamma^{2}
\left(-i\partial_{y}+ s_\perp \frac{x-x^{\prime}}{2l^{2}}\right)\right] \frac{e^{-\xi/2}}{2\pi l^{2}}
\sum\limits_{n=0}^\infty
\frac{L_{n}(\xi){\cal P}_{-}+L_{n-1}(\xi){\cal P}_{+}}{{\cal M}- 2n|e B |} .
\label{propagatorG-FT}
\end{equation}
Note that the ordering of the matrix factors in this expression is very important
because the expression in the square brackets does not commute with matrix ${\cal M}$.

The Fourier transform of the translation invariant part of propagator (\ref{propagatorG-FT}), 
\begin{equation}
\bar{G}(\omega,k^{3},\mathbf{k}) = \int d^2\mathbf{r}\, e^{-i (\mathbf{k} \cdot \mathbf{r})}
\bar{G}(\omega,k^{3};\mathbf{r}),
\end{equation}
can be evaluated by first performing the integration over the angle in the coordinate space. 
The integration is performed by making use of the following table integral: 
\begin{eqnarray}
\int\limits_{0}^{2\pi} e^{-ikr\cos(\phi-\phi_k)} d\phi &= & 2\pi J_{0}(kr),
\end{eqnarray}
where $J_{0}(x)$ is the Bessel function. Using also formula $7.421.1$ from Ref.~\cite{GR}, 
one gets
\begin{equation}
 \int_0^\infty
xe^{-\frac{1}{2}\alpha x^2}L_n\left(\frac{1}{2}\beta
x^2\right)J_0(xy)dx
=\frac{(\alpha-\beta)^n}{\alpha^{n+1}}e^{-\frac{1}{2\alpha}
y^2}L_n\left( \frac{\beta y^2}{2\alpha(\beta-\alpha)}\right),
\end{equation}
valid for $y>0$ and $\mbox{Re}\,\alpha>0$. The result is given by
\begin{equation}
\bar{G}(\omega,k^{3},\mathbf{k}) = ie^{-k^2 l^{2}}\sum_{n=0}^{\infty}
(-1)^nD_{n}(\omega,k^{3},\mathbf{k})\,\frac{1}{{\cal M}-2n|e B |},
\label{GDn-new}
\end{equation}
where $\mathbf{k}=(k^1,k^2)$ is the ``transverse momentum", $k^2=|\mathbf{k}|^2$
and the $n$th Landau level contribution is determined by
\begin{eqnarray}
D_{n}(\omega,k^{3},\mathbf{k}) = 2W\left[{\cal P}_{-}L_n\left(2 k^2 l^{2}\right)
-{\cal P}_{+}L_{n-1}\left(2 k^2 l^{2}\right)\right]
 + 4(\mathbf{k}\cdot\bm{\gamma}) L_{n-1}^1\left(2 k^2 l^{2}\right).
\label{Dn}
\end{eqnarray}
It is assumed that $L_{-1}^\alpha \equiv 0$. This could be compared with the standard
Dirac propagator for massive fermions in a constant magnetic field in Ref.~\cite{MC2}.
The last matrix factor in Eq.~(\ref{GDn-new}) can be rewritten in a more convenient
form as
\begin{equation}
\frac{1}{{\cal M}-2n|e B |} =
\frac{ (\omega+\mu-i\tilde{\mu}\gamma^0\gamma^1\gamma^2)^2
-(m+i\Delta\gamma^0\gamma^1\gamma^2)^2
+2i\gamma^1\gamma^2(\tilde{\mu}+ \Delta \gamma^0)k^{3}\gamma^3
-(k^{3})^2- 2n|e B |}{U_n},
\label{poles}
\end{equation}
where
\begin{equation}
U_n =\left[(\omega+\mu)^2 + \tilde{\mu}^2 - m^2 - \Delta^2 - (k^{3})^2 -
2n|e B |\right]^2- 4\left[(\tilde{\mu}\,(\omega+\mu) +
m\Delta)^2+(k^{3})^2(\Delta^2-\tilde{\mu}^2)\right].
\label{U_n}
\end{equation}
As follows from this representation, the fermion dispersion relations are determined
by the zeros of $U_n$, which is a fourth order polynomial in $\omega$ in a general
case.

\subsection{Fermion propagator in the coincidence limit, $u^\prime\to u$}

The coordinate space representation of the propagator reads
\begin{equation}
G(u,u^\prime)=e^{i\Phi(\mathbf{r},\mathbf{r}^\prime)}\int\frac{d\omega dk^{3} d^2\mathbf{k}}{(2\pi)^4}
e^{-i\omega (t-t^\prime)+ik^{3}(z-z^\prime)+i\mathbf{k}\cdot(\mathbf{r}-\mathbf{r}^\prime)}
\bar{G}(\omega,k^{3},\mathbf{k}).
\label{AppGuu}
\end{equation}
As seen from Eq.~(\ref{gap}) in Sec.~\ref{GapEquation}, only the full fermion propagator $G(u,u)$ 
in the coincidence limit, $u^\prime\to u$, enters the mean-field gap equation. By making use of the  
results in the previous subsection, we obtain the following expression for the corresponding propagator:
\begin{equation}
G(u,u) = \int\frac{d\omega dk^{3} d^2\mathbf{k}}{(2\pi)^4}
\bar{G}(\omega,k^{3},\mathbf{k})
= \frac{i}{2\pi l^2}\sum_{n=0}^{\infty} \int\frac{d\omega d k^{3}}{(2\pi)^2}
\frac{{\cal K}_{n}^{-}{\cal P}_{-}+{\cal K}_{n}^{+}{\cal P}_{+}\theta(n-1)}{U_n},
\label{Guu}
\end{equation}
where $\theta(n-1)=1$ for $n\geq 1$ and $\theta(n-1)=0$ for $n\leq 1$. We took into 
account that $\Phi(\mathbf{r},\mathbf{r}^\prime)=0$ at $\mathbf{r}^\prime=\mathbf{r}$, used the 
table integral 7.414 (8) in Ref.~\cite{GR},
\begin{equation}
\int_{0}^{\infty} e^{-st} t^\alpha L_n^\alpha(t) dt = \frac{\Gamma(\alpha+n+1)(s-1)^n}{s^{n+1}} , 
\label{integral7.414}
\end{equation}
(valid when $\mbox{Re}\,\alpha>-1$ and $\mbox{Re}\,s>0$)
and introduced the following shorthand notation:
\begin{eqnarray}
{\cal K}_{n}^{\pm}&=& \left[(\omega+\mu \mp s_{\perp}\Delta)\gamma^0
\pm s_{\perp}\tilde\mu
+ m  -k^{3}\gamma^3\right]\Big\{
(\omega+\mu)^2 + \tilde{\mu}^2 -  m ^2 - \Delta^2 - (k^{3})^2 - 2n|eB_{\perp}| \nonumber \\
&& \mp 2 s_{\perp}\left[\tilde\mu (\omega+\mu)+\Delta m \right]\gamma^0
\pm 2 s_{\perp} (\tilde\mu + \Delta\gamma^0)k^{3} \gamma^3\Big\}.
\label{K_n^pm}
\end{eqnarray}
Note that the first factor in the square brackets originates from matrix $W$, see Eqs.(\ref{matrices-MW}),
(\ref{GDn-new}) and (\ref{Dn}). 

In the rest of this Appendix, we consider only the case with the vanishing anomalous magnetic
moment ($\tilde\mu=0$), which is sufficient for the mean-field analysis in the NJL model. The 
expressions for ${\cal K}_n$ and $U_n$ in this special case are quoted in 
Eqs.~(\ref{K_n^pm_text}) and (\ref{U_n_text}) in the main text. Then, as follows from 
Eq.~(\ref{Guu}), the propagator has the following general structure:
\begin{equation}
G(u,u)=G_0^{-}{\cal P}^{-}+\sum_{n=1}^{\infty}\left(G_{n}^{-}{\cal P}^{-}+G_{n}^{+}{\cal P}^{+}\right).
\label{Guu-Gn}
\end{equation}
By substituting $\tilde\mu=0$ and replacing the integration over $\omega$ by the Matsubara sum 
according to the prescription
\begin{equation}
\int\frac{d \omega}{2\pi}(\cdots) \to i T\sum_{m^\prime=-\infty}^{\infty} (\cdots),
\label{finiteTprescription}
\end{equation}
with $\omega\to i\omega_{m^\prime}= i \pi T(2m^\prime+1)$, we derive separate $n$th Landau level contributions, 
\begin{eqnarray}
G_n^{\pm} &=& -\frac{T}{2\pi l^2}\int_{-\infty}^{\infty}\frac{d k^{3}}{2\pi} \sum_{m^\prime=-\infty}^{\infty}
\left.\frac{{\cal K}_{n}^{\pm}}{U_{n}}\right|_{\omega\to i\omega_{m^\prime}} \nonumber\\
&=&\frac{T}{4\pi l^2}\int_{-\infty}^{\infty}\frac{d k^{3}}{2\pi} \sum_{m^\prime=-\infty}^{\infty}
\left(\gamma^0\pm \frac{ m-k^{3}\gamma^3 }{\sqrt{ m ^2 +(k^{3})^2}}\right)
\frac{i\omega_{m^\prime}+\mu \pm \left(\sqrt{ m ^2 +(k^{3})^2}- s_{\perp} \Delta\right)} 
{(\omega_{m^\prime}-i\mu)^2 + \left(E_{k^{3},n}^{-}\right)^2}\nonumber\\
&+&\frac{T}{4\pi l^2}\int_{-\infty}^{\infty}\frac{d k^{3}}{2\pi} \sum_{m^\prime=-\infty}^{\infty}
\left(\gamma^0\mp \frac{ m-k^{3}\gamma^3 }{\sqrt{ m ^2 +(k^{3})^2}}\right)
\frac{i\omega_{m^\prime}+\mu \mp  \left(\sqrt{ m ^2 +(k^{3})^2}+ s_{\perp} \Delta\right)}
 {(\omega_{m^\prime}-i\mu)^2 + \left(E_{k^{3},n}^{+}\right)^2},
\end{eqnarray}
where
$E_{k^{3},n}^{\pm}\equiv \sqrt{\left(s_{\perp}\Delta\pm \sqrt{m^2+(k^{3})^2} \right)^2 + 2n|e B |}$.
We use the following table sums:
\begin{eqnarray}
T\sum_{{m^\prime}=-\infty}^{\infty}\frac{b}{(\omega_{m^\prime}-i\mu)^2 + b^2}  &=& 
\frac{1-n_F(b+\mu)-n_F(b-\mu)}{2},
\label{tablesum1}\\
T\sum_{{m^\prime}=-\infty}^{\infty}\frac{i\omega_{m^\prime} + \mu}{(\omega_{m^\prime} -i \mu)^2 + b^2 } 
&=&\frac{n_{F}(b+\mu)-n_{F}(b-\mu)}{2},
\label{tablesum2}
\end{eqnarray}
where $n_{F}(x) =1/(e^x+1)$ is the Fermi-Dirac distribution function, and finally obtain
\begin{eqnarray}
G_n^{\pm} &=&
\frac{1}{8\pi l^2}\int_{-\infty}^{\infty}\frac{d k^{3}}{2\pi}
\left(\gamma^0\pm \frac{ m }{\sqrt{ m ^2 +(k^{3})^2}}\right)
\left[n_{F}\left(E_{k^{3},n}^{-}+\mu \right)-n_{F}\left(E_{k^{3},n}^{-}-\mu \right) \right]
\nonumber\\
&+&\frac{1}{8\pi l^2}\int_{-\infty}^{\infty}\frac{d k^{3}}{2\pi}
\left(\gamma^0\mp\frac{ m }{\sqrt{ m ^2 +(k^{3})^2}}\right)
\left[n_{F}\left(E_{k^{3},n}^{+}+\mu \right)-n_{F}\left(E_{k^{3},n}^{+}-\mu \right) \right]
\nonumber\\
&\pm&\frac{1}{8\pi l^2}\int_{-\infty}^{\infty}\frac{d k^{3}}{2\pi}
\left( \gamma^0 \pm \frac{ m }{\sqrt{ m ^2 +(k^{3})^2}} \right)
\frac{\sqrt{ m ^2 +(k^{3})^2}-s_{\perp} \Delta}{E_{k^{3},n}^{-}}
\left[1-n_{F}\left(E_{k^{3},n}^{-}+\mu \right) -n_{F}\left(E_{k^{3},n}^{-}-\mu \right)\right]
\nonumber\\
&\mp&\frac{1}{8\pi l^2}\int_{-\infty}^{\infty}\frac{d k^{3}}{2\pi}
\left( \gamma^0 \mp \frac{ m }{\sqrt{ m ^2 +(k^{3})^2}} \right)
\frac{\sqrt{ m ^2 +(k^{3})^2}+s_{\perp} \Delta}{E_{k^{3},n}^{+}}
\left[1-n_{F}\left(E_{k^{3},n}^{+}+\mu \right)-n_{F}\left(E_{k^{3},n}^{+}-\mu \right) \right].
\label{AppGn}
\end{eqnarray}
The terms odd in $k^{3}$ were dropped because they vanish
after the integration over $k^{3}$ is performed.

In the case of the lowest Landau level ($n=0$), the result can be rewritten in a more convenient 
form as
\begin{eqnarray}
G_0^{-} &=& \frac{1}{4\pi l^2}\int_{-\infty}^{\infty}\frac{d k^{3}}{2\pi}
\frac{ m }{\sqrt{ m ^2 +(k^{3})^2}}
\left[1-n_{F}\left(\sqrt{ m ^2 +(k^{3})^2}-s_{\perp} \Delta+\mu\right)-n_{F}
\left(\sqrt{ m ^2 +(k^{3})^2}+s_{\perp} \Delta -\mu \right) \right]
\nonumber\\
&+&\frac{1}{4\pi l^2}\gamma^0 \int_{-\infty}^{\infty}\frac{d k^{3}}{2\pi}
\left[n_{F}\left(\sqrt{ m ^2 +(k^{3})^2}-s_{\perp} \Delta+\mu\right)-n_{F}
\left(\sqrt{ m ^2 +(k^{3})^2}+s_{\perp} \Delta -\mu \right) \right].
\label{AppG0}
\end{eqnarray}
We notice that functions $G_n^\pm$ contain only two different Dirac structures: $\gamma^0$ 
and the unit matrix. Taking this into account, we conclude that the fermion propagator in the 
coincidence limit, $u^\prime\to u$, is given in terms of just four independent Dirac structures:
\begin{equation}
G(u,u)=-\frac{1}{4}\left[\gamma^0{\cal A}+{\cal B}+ i\gamma^1\gamma^2{\cal C}+ \gamma^3\gamma^5{\cal D}\right].
\label{Guu-ABCD}
\end{equation}
The explicit expressions for functions $ {\cal A}$, $ {\cal B}$, $ {\cal C}$  and $ {\cal D}$ follow 
from the results in Eqs.~(\ref{AppGn}) and (\ref{AppG0}), namely
\begin{eqnarray}
 {\cal A} &=& -\frac{1}{(2\pi l)^2} \int_{-\infty}^{\infty} dk^{3}
\left[
 n_{F}\left(\sqrt{ m ^2 +(k^{3})^2}-s_{\perp} \Delta +\mu\right)
-n_{F}\left(\sqrt{ m ^2 +(k^{3})^2}+s_{\perp} \Delta -\mu \right) 
\right]
\nonumber\\
&& -\frac{1}{(2\pi l)^2} \sum_{n=1}^{\infty}\int_{-\infty}^{\infty} dk^{3}\left[
 n_{F}\left(E_{k^{3},n}^{-}+\mu \right)
-n_{F}\left(E_{k^{3},n}^{-}-\mu \right)
+n_{F}\left(E_{k^{3},n}^{+}+\mu \right)
-n_{F}\left(E_{k^{3},n}^{+}-\mu \right)
\right] , \label{functionA}
\\
{\cal B}&=&  - \frac{m}{(2\pi l)^2}\int_{-\infty}^{\infty}
\frac{ d k^{3}}{\sqrt{ m ^2 +(k^{3})^2}}
\left[1-n_{F}\left(\sqrt{ m ^2 +(k^{3})^2}-s_{\perp} \Delta+\mu\right)
-n_{F}\left(\sqrt{ m ^2 +(k^{3})^2}+s_{\perp} \Delta -\mu \right) \right]
\nonumber\\
&&-\frac{m}{(2\pi l)^2}  \sum_{n=1}^{\infty}\int_{-\infty}^{\infty}
\frac{dk^{3}}{\sqrt{ m ^2 +(k^{3})^2}}\left\{
\frac{\sqrt{ m ^2 +(k^{3})^2}-s_{\perp} \Delta}{E_{k^{3},n}^{-}}
\left[1
-n_{F}\left(E_{k^{3},n}^{-}+\mu \right)
-n_{F}\left(E_{k^{3},n}^{-}-\mu \right)
\right]+
\right.
\nonumber\\
&&\left.\hspace{1.5in}
+\frac{\sqrt{ m ^2 +(k^{3})^2}+s_{\perp} \Delta}{E_{k^{3},n}^{+}}
\left[1
-n_{F}\left(E_{k^{3},n}^{+}+\mu \right)
-n_{F}\left(E_{k^{3},n}^{+}-\mu \right)
\right]
\right\}\label{functionB}
,\\
{\cal C}&=& \frac{m s_{\perp}}{(2\pi l)^2}\int_{-\infty}^{\infty}
\frac{ d k^{3}}{\sqrt{ m ^2 +(k^{3})^2}}
\left[1-n_{F}\left(\sqrt{ m ^2 +(k^{3})^2}-s_{\perp} \Delta+\mu\right)
-n_{F}\left(\sqrt{ m ^2 +(k^{3})^2}+s_{\perp} \Delta -\mu \right) \right]
\nonumber\\
&+&\frac{m s_{\perp}}{(2\pi l)^2} \sum_{n=1}^{\infty}\int_{-\infty}^{\infty} \frac{dk^{3}}{\sqrt{ m ^2 +(k^{3})^2}}\left[
n_{F}\left(E_{k^{3},n}^{-}-\mu \right)-n_{F}\left(E_{k^{3},n}^{-}+\mu \right)
-n_{F}\left(E_{k^{3},n}^{+}-\mu \right)+n_{F}\left(E_{k^{3},n}^{+}+\mu \right)
\right]\label{functionC}
,\\
{\cal D} &=& \frac{ s_{\perp}}{(2\pi l)^2} \int_{-\infty}^{\infty} dk^{3}
\left[n_{F}\left(\sqrt{ m ^2 +(k^{3})^2}-s_{\perp} \Delta +\mu\right)
-n_{F}\left(\sqrt{ m ^2 +(k^{3})^2}+s_{\perp} \Delta -\mu \right) \right]
\nonumber\\
&&-\frac{ s_{\perp}}{(2\pi l)^2}  \sum_{n=1}^{\infty}\int_{-\infty}^{\infty} dk^{3}\left\{
\frac{\sqrt{ m ^2 +(k^{3})^2}-s_{\perp} \Delta}{E_{k^{3},n}^{-}}
\left[1
-n_{F}\left(E_{k^{3},n}^{-}+\mu \right)
-n_{F}\left(E_{k^{3},n}^{-}-\mu \right)
\right]
\right.
\nonumber\\
&&\left.\hspace{1in}
-\frac{\sqrt{ m ^2 +(k^{3})^2}+s_{\perp} \Delta}{E_{k^{3},n}^{+}}
\left[1
-n_{F}\left(E_{k^{3},n}^{+}+\mu \right)
-n_{F}\left(E_{k^{3},n}^{+}-\mu \right)
\right]
\right\}.
\label{functionD}
\end{eqnarray}
In order to clarify the physical meaning of these functions, it is useful to note their alternative 
definitions in terms of the following traces of the propagator:
\begin{eqnarray}
 {\cal A} &=&-\tr\left[\gamma^0G(u,u)\right]
 \equiv \langle \bar\psi \gamma^0 \psi\rangle = \langle j^0 \rangle, 
 \label{A-density}\\
{\cal B}&=& -\tr\left[G(u,u)\right]
\equiv \langle \bar\psi \psi\rangle,
\label{B-chi-condensate}\\
{\cal C}&=& -\tr\left[ i\gamma^1\gamma^2G(u,u)\right]
\equiv  \langle\bar\psi i\gamma^1\gamma^2\psi\rangle, 
\label{C-anomalous-moment}\\
{\cal D} &=&-\tr\left[\gamma^3\gamma^5G(u,u)\right]
\equiv \langle \bar\psi \gamma^3\gamma^5\psi\rangle=  \langle j_5^3 \rangle.
\label{D-axial-current}
\end{eqnarray}
Note that out of the four Dirac structures in the fermion propagator, only three of them $ {\cal A}$, $ {\cal B}$ 
and $ {\cal D}$ appear in the gap equation. Out of these latter, only two functions, i.e., ${\cal B}$ and ${\cal D}$, 
contain ultraviolet divergences. In Appendix~\ref{AppFunctionsABD} below, we analyze functions $ {\cal A}$, 
$ {\cal B}$ and $ {\cal D}$ at zero temperature using the proper time regularization.

\section{Function ${\cal A}$, ${\cal B}$ and ${\cal D}$ in proper time regularization}
\label{AppFunctionsABD}

\subsection{Function ${\cal A}$ at $T=0$}

As seen from Eq.~(\ref{A-density}), function ${\cal A}$ coincides with the baryon number density. 
As expected, the corresponding expression in Eq.~(\ref{functionA}) has no divergences. Moreover, 
at $T=0$, it can be given in a closed form in terms of elementary functions,
\begin{eqnarray}
{\cal A}|_{T=0} &=& \frac{\sign(\mu-s_{\perp}\Delta)}{2(\pi l)^2}\sqrt{(\mu-s_{\perp}\Delta)^2-m^2}
\theta\left(|\mu-s_{\perp}\Delta|-|m|\right) \nonumber\\
&+&
\frac{\sign(\mu)}{2(\pi l)^2} \sum_{n=1}^{N_B}
\sqrt{\left(\sqrt{\mu^2-2n|eB|}-|\Delta|\right)^2-m^2} \,
\sign\left(\sqrt{\mu^2-2n|eB|}-|\Delta|\right)
\theta\left[\left(\sqrt{\mu^2-2n|eB|}-|\Delta|\right)^2-m^2\right]
 \nonumber\\
&+&
\frac{\sign(\mu)}{2(\pi l)^2} \sum_{n=1}^{N_B}
\sqrt{\left(\sqrt{\mu^2-2n|eB|}+|\Delta|\right)^2-m^2} \,
\theta\left[\left(\sqrt{\mu^2-2n|eB|}+|\Delta|\right)^2-m^2\right],
\label{functionA@T=0}
\end{eqnarray}
where $N_B$ is the integer part of $\mu^2/(2|eB|)$.
At $m=0$, in particular, this becomes
\begin{equation}
{\cal A}|_{T=0,m=0} =\frac{ \mu-s_{\perp}\Delta }{2(\pi l)^2} +\frac{\sign(\mu)}{(\pi l)^2} \sum_{n=1}^{N_B}
\sqrt{\mu^2-2n|eB|} .
\label{functionA@T=0&m=0}
\end{equation}

\subsection{Function ${\cal B}$ at $T=0$}

As clear from Eq.~(\ref{B-chi-condensate}), function ${\cal B}$ coincides with the chiral condensate. We start by
studying the divergent part of this function. As can be seen from Eq.~(\ref{functionB}), all the divergences 
are independent not only of the temperature, but also of the chemical potential. Thus, at first we limit ourselves 
to the vacuum part ($\mu=0$) of the chiral condensate,
\begin{eqnarray}
{\cal B}_{T=0}^{\rm vac} &=& - \frac{m}{(2\pi l)^2}\int_{-\infty}^{\infty}
\frac{ d k^{3}}{\sqrt{ m ^2 +(k^{3})^2}}\left[
1-\theta\left(-\sqrt{ m ^2 +(k^{3})^2}+s_{\perp} \Delta \right)
-\theta\left(-\sqrt{ m ^2 +(k^{3})^2}-s_{\perp} \Delta\right)\right.
\nonumber\\
&&\hspace{1in}+\left. \sum_{n=1}^{\infty} \left(
\frac{\sqrt{ m ^2 +(k^{3})^2}-s_{\perp} \Delta}{E_{k^{3},n}^{-}}
+\frac{\sqrt{ m ^2 +(k^{3})^2}+s_{\perp} \Delta}{E_{k^{3},n}^{+}}
\right)\right].
\end{eqnarray}
By reintroducing integration over frequency, this expression can be equivalently rewritten as follows:
\begin{eqnarray}
{\cal B}_{T=0}^{\rm vac} &=& - \frac{m}{(2\pi l)^2} \int_{-\infty}^{\infty} \frac{ d k^{3}}{\sqrt{ m ^2 +(k^{3})^2}}
\sum_{n=0}^{\infty}\alpha_n\left(
\frac{\sqrt{ m ^2 +(k^{3})^2}-s_{\perp} \Delta}{E_{k^{3},n}^{-}}
+\frac{\sqrt{ m ^2 +(k^{3})^2}+s_{\perp} \Delta}{E_{k^{3},n}^{+}}
\right)
\nonumber\\
&=&  - \frac{m}{(2\pi l)^2} \int\frac{d\omega}{\pi}
\int_{-\infty}^{\infty} \frac{ d k}{\sqrt{ m ^2 +(k)^2}}
\sum_{n=0}^{\infty}\alpha_n\left(
\frac{\sqrt{m^2+k^2}-s_{\perp}\Delta}{\omega^2+\left(E_{k^{3},n}^{-}\right)^2 }
+\frac{\sqrt{m^2+k^2}+s_{\perp}\Delta}{\omega^2+\left(E_{k^{3},n}^{+}\right)^2 }
\right)
\nonumber\\
&=& - \frac{2m}{(2\pi l)^2} \int\frac{d\omega}{\pi}
\int_{-\infty}^{\infty}  d k
\sum_{n=0}^{\infty}\alpha_n
\frac{\omega^2+m^2+k^2 -\Delta^2+2n|eB|}{\left[\omega^2+(\sqrt{m^2+k^2}-s_{\perp}\Delta)^2+2n|eB|\right]
\left[\omega^2+(\sqrt{m^2+k^2}+s_{\perp}\Delta)^2+2n|eB|\right]},\nonumber
\\
\label{b4}
\end{eqnarray}
where, by definition, $ \alpha_{n} =1-\frac{1}{2}\delta_{n}^{0}$. The divergent expression on the right hand 
side can be regularized by making use of the proper time method. In contrast to the cutoff regularization 
used in the main text, this regularization is explicitly gauge invariant. 

In Eq.~(\ref{b4}), we introduce a proper time representation for each of the two factors in the denominator 
of the integrand and obtain the following result:
\begin{eqnarray}
{\cal B}_{T=0}^{\rm vac} &=& - \frac{2m}{(2\pi l)^2} \sum_{n=0}^{\infty}\alpha_n
\int\frac{d\omega}{\pi}  \int_{-\infty}^{\infty}  d k
\int_0^{\infty}ds_1\int_0^{\infty}ds_2\,\,
(\omega^2+m^2+k^2-\Delta^2+2n|eB|) e^{-2n(s_1+s_2)|eB|}
\nonumber\\
&&\times  e^{-(s_1+s_2)(\omega^2+m^2+k^2+\Delta^2)}  e^{2s_{\perp}\Delta(s_2-s_1)\sqrt{m^2+k^2} }.
\end{eqnarray}
Here it is convenient to introduce two new integration variables $s$ and $u$ instead of the proper 
times $s_1$ and $s_2$, which are related as follows:
\begin{equation}
s_1=\frac{s}{2}(1-u),\qquad s_2=\frac{s}{2}(1+u).
\label{new-variables}
\end{equation}
The integration over $s$ runs from $0$ to $\infty$, the integration over $u$ runs from $-1$ to $1$, and the 
Jacobian associated with the change of variables is $s/2$. Then, summing over the Landau quantum number 
$n$ and integrating over $\omega$ and $u$, we derive  
\begin{equation}
{\cal B}_{T=0}^{\rm vac} = -\frac{2m}{\sqrt{\pi}(4\pi l)^2}
\int_{-\infty}^{\infty}  d k
\int_{1/\Lambda^2}^{\infty}\frac{ds}{\sqrt{s}}\frac{\sinh(2s\Delta \sqrt{m^2+k^2})}{\Delta \sqrt{m^2+k^2}}
\left[\left(\frac{1}{2s}+m^2+k^2-\Delta^2\right) \coth(eBs)+\frac{eB}{\sinh^2(eBs)}\right]
 e^{-s(m^2+k^2+\Delta^2)}  .
 \label{b7}
\end{equation}
Note that in the last expression we introduced a finite proper-time cutoff at $s=1/\Lambda^2$ to 
regularize the ultraviolet divergences. It is easy to see that the divergent part at $\Lambda\to \infty$  
is independent of $\Delta$. Thus, after taking the limit $\Delta\to 0$ and integrating over $k$, we arrive 
at the following result:
\begin{eqnarray}
{\cal B}_{T=0}^{\rm vac} &=& -\frac{m}{2(2\pi l)^2}
\int_{1/\Lambda^2}^{\infty}ds \left[\left(\frac{1}{s}+m^2\right) \coth(eBs)+\frac{eB}{\sinh^2(eBs)}\right] e^{-s m^2 }  ,
\nonumber\\
&=&-\frac{m}{2(2\pi l)^2}\left[\coth\left(\frac{eB}{\Lambda^2}\right) e^{- (m/\Lambda)^2 } 
+\int_{1/\Lambda^2}^{\infty}\frac{ds}{s}\coth(eBs) \right],
 \label{b8}
\end{eqnarray}
where, after noting that $eB/\sinh^2(eBs)=-\frac{d}{ds}\coth(eBs) $, we were able to simplify the second term 
in the square brackets by integrating it by parts. Finally, we derive the result,
\begin{eqnarray}
{\cal B}_{T=0}^{\rm vac}|_{\Delta=0} &\simeq &  - \frac{m}{(2\pi l)^2}  \Big[ \coth\left(\frac{eB}{\Lambda^2}\right)  e^{-m^2/\Lambda^2}
+(\Lambda l)^2 e^{-m^2/\Lambda^2 } -(ml)^2 \mbox{E}_1(m^2/\Lambda^2)
 \nonumber\\
&&  + \ln\frac{1}{\pi (ml)^2}+2\ln\Gamma\left(1+\frac{(ml)^2}{2}\right) + (ml)^2 \left(1+\ln 2-\ln (ml)^2\right)\Big] \nonumber\\
&\simeq &   -\frac{2\Lambda^2 m}{(2\pi)^2}
- \frac{m}{(2\pi l)^2} \ln\frac{1}{\pi(m l)^2}
- \frac{m^3}{(2\pi)^2}\left[\ln(\Lambda l)^2+\ln 2-1\right]+O\left(\frac{m}{l^2(\Lambda l)^2}, m^5l^2\right).
\label{B-vacuum}
\end{eqnarray}
By making use of the following leading order approximation for the remaining proper time integration:
\begin{eqnarray}
\int_{\epsilon}^{\infty}\frac{d\tau}{\tau}e^{-\tau a^2 }\coth\tau
&\simeq & \int_{\epsilon}^{\infty}\frac{d\tau}{\tau^2}e^{-\tau a^2 }
+\int_{0}^{\infty}\frac{d\tau}{\tau^2}e^{-\tau a^2 }\left(\tau\coth\tau-1\right)+O(\epsilon)\nonumber\\
&= &  \frac{e^{-\epsilon a^2 }}{\epsilon}-a^2 \mbox{E}_1(\epsilon a^2)
+ \ln\frac{1}{\pi a^2}+2\ln\Gamma\left(1+\frac{a^2}{2}\right)
+  a^2 \left(1+\ln 2-\ln a^2\right)+O(\epsilon),
\end{eqnarray}
where $\mbox{E}_1(z)$ is the exponential integral function and we use the same identities for the generalized
Riemann zeta functions as in Ref.~\cite{MC2}. Note that $\mbox{E}_1(z)\simeq -\ln z -\gamma_E+z +O(z^2)$  for $z\to 0$.

The matter part of the function ${\cal B}$ at zero temperature reads:
\begin{eqnarray}
{\cal B}_{T=0}^{\rm matter}  &\equiv& {\cal B}_{T=0}  -{\cal B}_{T=0}^{\rm vac}\nonumber
\\
&=&\frac{m}{2(\pi l)^2} \left[\ln\frac{|\mu-s_{\perp}\Delta|
+\sqrt{(\mu-s_{\perp}\Delta)^2-m^2}}{|m|}\theta\left(|\mu-s_{\perp}\Delta|-|m|\right)
-\ln\frac{|\Delta|+\sqrt{\Delta^2-m^2}}{|m|}\theta\left(|\Delta|-|m|\right)
\right] \nonumber
\\
&+&\frac{m}{(2\pi l)^2}\sum_{n=1}^{\infty}
\int_{-\infty}^{\infty} \frac{dk^{3}}{\sqrt{m^2+(k^{3})^2}}
\left[\frac{\sqrt{ m ^2 +(k^{3})^2}-s_{\perp} \Delta}{E_{k^{3},n}^{-}}\theta\left(|\mu|-E_{k^{3},n}^{-}\right)
+\frac{\sqrt{ m ^2 +(k^{3})^2}+s_{\perp} \Delta}{E_{k^{3},n}^{+}}\theta\left(|\mu|-E_{k^{3},n}^{+}\right)
\right] .\nonumber
\\
\end{eqnarray}
In the limit $\Delta\to 0$, the result can be presented in an analytical form,
\begin{equation}
{\cal B}_{T=0}^{\rm matter}|_{\Delta=0} =
\frac{m}{2(\pi l)^2} \ln\frac{|\mu|+\sqrt{\mu^2-m^2}}{|m|}\theta\left(|\mu|-|m|\right)
+\frac{m}{(\pi l)^2}\sum_{n=1}^{\infty}
\ln\frac{|\mu|+\sqrt{\mu^2-m^2-2n|eB|}}{\sqrt{m^2+2n|eB|}}\theta\left(|\mu|-\sqrt{m^2+2n|eB|}\right).
\label{B-matter}
\end{equation}

\subsection{Function ${\cal D}$ at $T=0$}

The vacuum part ($\mu=0$) of the function ${\cal D}$ 
in Eq.~(\ref{functionD}) at zero temperature is given by
\begin{eqnarray}
{\cal D}_{T=0}^{\rm vac} &=& \frac{ s_{\perp}}{(2\pi l)^2} \int_{-\infty}^{\infty} dk^{3}
\left[\theta\left(-\sqrt{ m ^2 +(k^{3})^2}+s_{\perp} \Delta \right)
-\theta\left(-\sqrt{ m ^2 +(k^{3})^2}-s_{\perp} \Delta\right) \right]
\nonumber\\
&&-\frac{ s_{\perp}}{(2\pi l)^2}  \sum_{n=1}^{\infty}\int_{-\infty}^{\infty} dk^{3}\left(
\frac{\sqrt{ m ^2 +(k^{3})^2}-s_{\perp} \Delta}{E_{k^{3},n}^{-}}
-\frac{\sqrt{ m ^2 +(k^{3})^2}+s_{\perp} \Delta}{E_{k^{3},n}^{+}}
\right)\nonumber\\
&=& -\frac{ s_{\perp}}{(2\pi l)^2}  \int_{-\infty}^{\infty} dk^{3}\left[
\frac{1}{2}\frac{\sqrt{ m ^2 +(k^{3})^2}-s_{\perp} \Delta}{\sqrt{\sqrt{ (m ^2 +(k^{3})^2}-s_{\perp} \Delta)^2}}
-\frac{1}{2}\frac{\sqrt{ m ^2 +(k^{3})^2}+s_{\perp} \Delta}{\sqrt{\sqrt{ ( m ^2 +(k^{3})^2}+s_{\perp} \Delta)^2}}
\right.\nonumber\\
&&\left. \hspace{1in}
+\sum_{n=1}^{\infty}\left(
\frac{\sqrt{ m ^2 +(k^{3})^2}-s_{\perp} \Delta}{E_{k^{3},n}^{-}}
-\frac{\sqrt{ m ^2 +(k^{3})^2}+s_{\perp} \Delta}{E_{k^{3},n}^{+}}
\right)
\right]\nonumber\\
&=& -\frac{ s_{\perp}}{(2\pi l)^2}\int
\frac{d\omega}{\pi}\int_{-\infty}^{\infty}dk\,\sum_{n=0}^{\infty}\alpha_n\,
\left(\frac{\sqrt{m^2+k^2}-s_{\perp}\Delta}{\omega^2+(\sqrt{m^2+k^2}-s_{\perp}\Delta)^2+2neB}
-\frac{\sqrt{m^2+k^2}+s_{\perp}\Delta}{\omega^2+(\sqrt{m^2+k^2}+s_{\perp}\Delta)^2+2neB} \right)
\nonumber\\
&=&- \frac{2\Delta}{(2\pi l)^2}\int
\frac{d\omega}{\pi}\int_{-\infty}^{\infty}dk\,\sum_{n=0}^{\infty}\alpha_n\,
\frac{m^2+k^2-\omega^2-\Delta^2-2neB}{[\omega^2+(\sqrt{m^2+k^2}-s_{\perp}\Delta)^2+2neB]
\,[\omega^2+(\sqrt{m^2+k^2}+s_{\perp}\Delta)^2+2neB]}
\nonumber\\
&=&-\frac{2\Delta}{(2\pi l)^2}\sum_{n=0}^{\infty}\alpha_n\int
\frac{d\omega}{\pi}\int_{-\infty}^{+\infty}dk\int_0^{\infty}ds_1\int_0^{\infty}ds_2\,\,
(m^2+k^2-\omega^2-\Delta^2-2neB)e^{-2n(s_1+s_2)eB}
\nonumber\\
&&\times  e^{-(s_1+s_2)(\omega^2+m^2+k^2+\Delta^2)}  e^{2s_{\perp}\Delta(s_2-s_1)\sqrt{m^2+k^2} }.
\end{eqnarray}
We integrate over $\omega$, change the proper-time variables as in Eq.~(\ref{new-variables}), and 
obtain
\begin{eqnarray}
{\cal D}_{T=0}^{\rm vac}  &=& -\frac{1}{2\sqrt{\pi}(2\pi l)^2}
\int_{0}^{\infty}\frac{ds}{\sqrt{s} }\,
\int_{-\infty}^{+\infty}\frac{\sinh(2s \Delta\sqrt{m^2+k^2} )
dk }{ \sqrt{m^2+k^2} }
\left[ \left(m^2+k^2-\frac{1}{2s}-\Delta^2\right)\coth(eBs)-\frac{eB}{\sinh^2(eBs)}\right]
\nonumber\\
&&\times \,e^{-s( m^2+k^2+\Delta^2) }\,.
\end{eqnarray}
In the limit of $m\to 0$, in particular, this can be calculated analytically,
\begin{eqnarray}
{\cal D}_{T=0}^{\rm vac}|_{m\to 0}&=&
- \frac{\sqrt{\pi}}{2(2\pi l)^2}
\int_{0}^{\infty}\frac{ds}{\sqrt{s} }\,e^{-s \Delta^2 }
\mbox{erfi}(\sqrt{s} \Delta )
\left[ \coth(eBs)\left(-\frac{1}{2s}-\Delta^2\right)+\frac{d}{ds}\coth(eBs)\right] \nonumber\\
&-& \frac{\Delta}{2 (2\pi l)^2}
\int_{0}^{\infty}\frac{ds}{s} \coth(eBs) 
= \frac{\sqrt{\pi}}{2(2\pi l)^2}
\frac{e^{-s \Delta^2 }}{\sqrt{s} }\,
\mbox{erfi}(\sqrt{s} \Delta )\coth(eBs) \Bigg|_{s=1/\Lambda^2} \nonumber\\
&=& \frac{\sqrt{\pi}\Delta}{2(2\pi l)^2} e^{-(\Delta/\Lambda)^2 }\,\frac{\Lambda}{\Delta} 
\mbox{erfi}\left(\frac{\Delta}{\Lambda} \right)\coth\left(\frac{eB}{\Lambda^2} \right)
\simeq \frac{\Delta\Lambda^2}{(2\pi)^2}-\frac{2\Delta^3}{3(2\pi)^2}
+O\left(\frac{\Delta}{l^4\Lambda^2},\frac{\Delta^5}{\Lambda^2}\right),
\label{D-vacuum}
\end{eqnarray}
where we first integrated over $u$ and then integrated over $k$ by using following table integrals: 
\begin{eqnarray}
\int_{-\infty}^{+\infty}dk \frac{\sinh(2ak)}{k} e^{-sk^2} &=& \pi\,\mbox{erfi}\left(\frac{a}{\sqrt{s}}\right),\\
\int_{-\infty}^{+\infty} dk\, k\sinh(2ak)e^{-sk^2} &=& \frac{\sqrt{\pi}}{s^{3/2}}ae^{\frac{a^2}{s}} =-\pi\,\frac{d}{ds} \mbox{erfi}\left(\frac{a}{\sqrt{s}}\right),
\end{eqnarray}
where $\mbox{erfi}(a) \equiv -i\mbox{erf}(ia)$
is the imaginary error function, which has the following asymptotes:
\begin{eqnarray}
\mbox{erfi}(a) &\simeq & \frac{2a}{\sqrt{\pi}}\left[1+\frac{a^2}{3}+\frac{a^4}{10}+\frac{a^6}{42}+O\left(a^8\right)\right] ,\quad\mbox{for}\quad a\to 0,\\
\mbox{erfi}(a) &\simeq & \frac{e^{a^2}}{\sqrt{\pi}a}\left[1+\frac{1}{2a^2}+\frac{3}{4a^4}+\frac{15}{8a^6}+O\left(\frac{1}{a^8}\right)\right] ,\quad\mbox{for}\quad a\to \infty.
\end{eqnarray}

The matter part of the function ${\cal D}$ at zero temperature reads:
\begin{eqnarray}
{\cal D}_{T=0}^{\rm matter} &\equiv&{\cal D}_{T=0} - {\cal D}_{T=0}^{\rm vac}\nonumber\\
&=&-\frac{2 s_{\perp}\sign(\mu-s_{\perp}\Delta)}{(2\pi l)^2}\sqrt{(\mu-s_{\perp}\Delta)^2-m^2}
\theta\left(|\mu-s_{\perp}\Delta|-|m|\right)
-\frac{2 \sign(\Delta)}{(2\pi l)^2}\sqrt{ \Delta^2-m^2}
\theta\left(|\Delta|-|m|\right) \nonumber\\
&&+\frac{s_{\perp}}{(2\pi l)^2}\sum_{n=1}^{\infty}
\int_{-\infty}^{\infty} dk^{3}
\left[
 \frac{\sqrt{ m ^2 +(k^{3})^2}-s_{\perp} \Delta}{E_{k^{3},n}^{-}}\theta\left(|\mu|-E_{k^{3},n}^{-}\right)
-\frac{\sqrt{ m ^2 +(k^{3})^2}+s_{\perp} \Delta}{E_{k^{3},n}^{+}}\theta\left(|\mu|-E_{k^{3},n}^{+}\right)
\right]\nonumber\\
&\to& -\frac{2 s_{\perp}\mu}{(2\pi l)^2} \quad \mbox{for} \quad m\to 0.
\label{D-matter}
\end{eqnarray}

\section{Gap equation}
\label{AppGapEq}

In order to derive the gap equation, it is convenient to use the formalism of the effective 
action for composite operators \cite{potential}. In the mean-field approximation, the
corresponding effective action $\Gamma$ takes the following form:
\begin{eqnarray}
\Gamma(G)={-i}\,\mbox{Tr}\left[\mbox{Ln} G^{-1} +S^{-1}G-1\right]
&+&\frac{ G_{int}}{2}\int dt \int d^{3}\mathbf{r} \left\{
\left(\mbox{tr}\left[G(u,u)\right]\right)^{2}
-\left(\mbox{tr}\left[\gamma^{5}G(u,u)\right]\right)^{2}
\right.\nonumber\\
&-&\left.\mbox{tr}\left[G(u,u)G(u,u)\right]
+\mbox{tr}\left[\gamma^{5}G(u,u)\gamma^{5}G(u,u)\right]\right\},
\label{potential}
\end{eqnarray}
where $u=(t,\mathbf{r})$. The diagrammatic form of the equation is shown in Fig.~\ref{fig-eff-action}.
\begin{figure}[ht]
\begin{center}
\includegraphics[width=.65\textwidth]{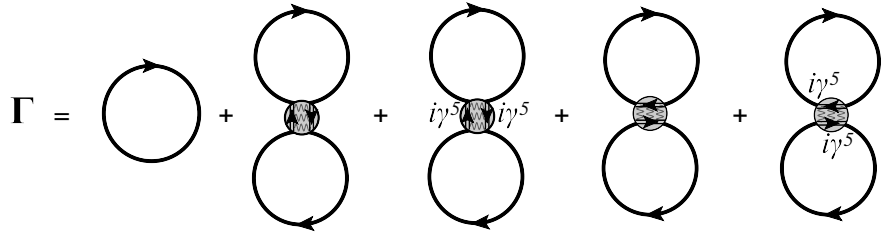}
\caption{Diagrammatic form of the expression for the effective action in the 
Hartree-Fock mean-field approximation.} 
\label{fig-eff-action}
\end{center}
\end{figure}
The trace, the logarithm, and the product $S^{-1}G$ are taken in the functional sense. 
The gap equation is obtained by requiring that the full fermion propagator $G$ corresponds 
to the variational extremum of the effective action, $\delta \Gamma/\delta G=0$, the 
explicit form of which reads
\begin{equation}
G^{-1}(u,u^\prime)  = S^{-1}(u,u^\prime)
- i G_{\rm int} \left\{ G(u,u) -  \gamma^5 G(u,u) \gamma^5 
-\mbox{tr}[G(u,u)] +  \gamma^5\, \mbox{tr}[\gamma^5G(u,u)]\right\}
\delta^{4}(u- u^\prime). 
\label{gap-App}
\end{equation}
By making use of the ansatz (\ref{ginverse}) for the full fermion propagator, this gap equation 
takes the following form:
\begin{equation}
(\mu-\mu_0)\gamma^0 +i\gamma^1\gamma^2 \tilde{\mu}
+i\Delta\gamma^0\gamma^1\gamma^2 -m+m_0 = -\frac{1}{2}G_{\rm int}
\left[\gamma^0  {\cal A}+\gamma^3\gamma^5 {\cal D}\right] +G_{\rm int}  {\cal B}.
\label{gap-more}
\end{equation}
Note that function ${\cal C}$ (anomalous magnetic moment) does not contribute to the right hand
side of the equation. This is in accordance with the statement in the main text that the anomalous
magnetic moment vanishes in the mean-field approximation used in our analysis of the NJL model.
This matrix equation is equivalent to the following set of algebraic equations:
\begin{eqnarray}
\mu &=&\mu_0 -\frac{1}{2}G_{\rm int}  {\cal A}  ,
\label{gap-mu}  \\
\Delta &=& -\frac{1}{2}G_{\rm int} {\cal D}    ,
\label{gap-Delta} \\
m &=& m_0 - G_{\rm int}  {\cal B}    ,
\label{gap-m}
\end{eqnarray}
together with $ \tilde{\mu}=0$.
An alternative form of the same equations is given in terms of the baryon number density,
the chiral condensate and the axial current density,
\begin{eqnarray}
\mu &=& \mu_0 -\frac{1}{2}G_{\rm int} \langle j^0\rangle ,
\label{gap-mu-x}  \\
\Delta &=& -\frac{1}{2}G_{\rm int} \langle j_5^3\rangle ,
\label{gap-Delta-x} \\
m &=&  m_0 - G_{\rm int} \langle \bar\psi \psi\rangle .
\label{gap-m-x}
\end{eqnarray}

\section{Free Energy Density}
\label{free-energy}

In this Appendix, we derive the expression for the free energy density $\Omega$. It can be given in
terms of the effective action $\Gamma$ evaluated at the solution to the gap equation,  
$\Omega = -\Gamma/{\cal T}V$, where ${\cal T}V$ is a space-time volume.  Taking into account 
the general form of the gap equation, the corresponding expression becomes
\begin{equation}
\Gamma= -i\,\mbox{Tr}\left[\mbox{Ln} G^{-1}
+\frac{1}{2}\left(S^{-1}G-1\right)\right].
\label{omega}
\end{equation}
By making use of the following Fourier transform of the Green's function
$G(u,u^\prime)$:
\begin{equation}
G(u,u^{\prime})=\int\limits_{-\infty}^{\infty}\frac{d\omega}{2\pi}\,
e^{-i\omega(t-t^\prime)}G(\omega;\mathbf{r},
\mathbf{r}^\prime),
\end{equation}
we rewrite the effective action $\Gamma$ as
\begin{equation}
\Gamma =
-i\,{\cal T}\int\limits_{-\infty}^{\infty}\frac{d\omega}{2\pi}\mbox{Tr}\left[\ln
G^{-1}(\omega)
+\frac{1}{2}\left(S^{-1}(\omega)G(\omega)-1\right)\right].
\label{effpot-integral-in-omega}
\end{equation}
Then, by following the approach of Ref.~\cite{GGMS2008}, we obtain the expression for the
free energy density,
\begin{eqnarray}
\Omega &=& -\int\limits_{-\infty}^{\infty}\frac{d\omega}{4\pi}\int
\frac{d^{3}k}{(2\pi)^{3}}
\mbox{tr}\left\{\left[(\omega-\mu_0)\gamma^{0}+(\mathbf{k}\cdot\bm{\gamma})+k^{3}\gamma^3\right]
\bar{G}(\omega,k^{3},\mathbf{k})+i\right\}.
\label{Omega}
\end{eqnarray}
The propagator $\bar{G}_{s}(\omega,k^{3},\mathbf{k})$ is given in
Eq.~(\ref{GDn-new}) in Appendix~\ref{AppPropagator}. By making use of its
explicit form and the table integral in Eq.~(\ref{integral7.414}), we can calculate 
the following three integrals that contribute to the free energy density,
\begin{eqnarray}
\int\frac{dk^{3} d^{2}\mathbf{k}}{(2\pi)^{3}}(\omega-\mu_0)\gamma^{0}\bar{G}_{s}(\omega,\mathbf{k})
&=&\frac{i}{(2\pi l)^2}\sum\limits_{n=0}^{\infty}\int{dk^{3}}
\frac{(\omega-\mu_0)\gamma^0  W \left( a_n-b\gamma^0 \right)\left[{\cal P}_{-}+{\cal P}_{+}\theta(n-1)\right]} {U_n},\label{int-1}\\
\int\frac{dk^{3} d^{2}\mathbf{k}}{(2\pi)^{3}}(\mathbf{k}\cdot\bm{\gamma})
\bar{G}_{s}(\omega,\mathbf{k}) &=&
 \frac{i}{(2\pi l)^2}\sum\limits_{n=0}^{\infty}\int{dk^{3}}
\frac{\left( a_n-b\gamma^0 \right)2n|e B |\left[{\cal P}_{-}+{\cal P}_{+}\theta(n-1)\right]}{U_n},\label{int-2}\\
\int\frac{dk^{3} d^{2}\mathbf{k}}{(2\pi)^{3}}k^{3}\gamma^3
\bar{G}_{s}(\omega,\mathbf{k}) &=& \frac{i}{(2\pi l)^2}\sum\limits_{n=0}^{\infty}
\int{dk^{3}}\frac{k^{3}\gamma^3
W \left( a_n-b\gamma^0 -(c+ d\gamma^0)k^{3}\gamma^3\right)
\left[{\cal P}_{-}+{\cal P}_{+}\theta(n-1)\right]} {U_n},
\label{int-3}
\end{eqnarray}
where $a_n$, $b$, $c$ and $d$ are mutually commuting functions, defined by the
following expressions:
\begin{eqnarray}
a_n &=& (\omega+\mu)^2 + \tilde{\mu}^2 - m^2 - \Delta^2 - (k^3)^2 - 2n|e B |,\\
b &=& 2i\gamma^1\gamma^2[(\omega+\mu)\tilde{\mu} + m\Delta],\\
c &=& -2i\gamma^1\gamma^2 \tilde{\mu},\\
d &=& -2i\gamma^1\gamma^2 \Delta.
\label{auxiliary}
\end{eqnarray}
[Note that the factor $\theta(n-1)$ in Eq.~(\ref{int-2}) is added for convenience;
it is optional because the result is proportional to $n$ and the $n=0$ term is
vanishing anyway.]

By dropping an infinite divergent term which is independent of the
physical parameters, from Eq.~(\ref{Omega}) we derive
the following expression for the free energy density:
\begin{eqnarray}
\Omega &=& -\frac{i}{(2\pi)^3l^2}\int\frac{d\omega dk^{3}}{U_0}
\Bigg\{
\left[(\omega+\mu)^2 + \tilde{\mu}^2 - m^2 - \Delta^2 - (k^{3})^2\right]
\left[(\omega-\mu_0)(\omega+\mu+s_{\perp} \Delta)+(k^{3})^2\right]
\nonumber\\
&&+2(\omega-\mu_0)(s_{\perp}  m -\tilde\mu)\left[\tilde\mu(\omega+\mu)+\Delta m \right]
+2s_{\perp} \tilde\mu (k^{3})^2( m -s_{\perp} \tilde\mu)+2s_{\perp} \Delta (k^{3})^2(\omega+\mu+s_{\perp} \Delta)
\Bigg\}
\nonumber\\
&-& \frac{2i}{(2\pi)^3l^2}\sum\limits_{n=1}^{\infty}
\int\frac{d\omega dk^{3}}{U_n} \Bigg\{
\left[(\omega+\mu)^2 + \tilde{\mu}^2 - m^2 - \Delta^2 - (k^{3})^2-2n|e B |\right]
\left[(\omega-\mu_0)(\omega+\mu)+(k^{3})^2+2n|e B |\right]
\nonumber\\
&&
-2(\omega-\mu_0)\tilde\mu \left[\tilde\mu(\omega+\mu)+ \Delta m \right]
+2(k^{3})^2\left(\Delta^2-\tilde\mu^2\right)
\Bigg\}.
\end{eqnarray}
Taking $\tilde\mu=0$, we have
\begin{eqnarray}
\Omega &=& -\frac{i}{2(2\pi)^3l^2}\int d\omega dk^{3}
\left[\frac{(\omega-\mu_0)(\omega+\mu+s_{\perp} \Delta)+(k^{3})^2}{(\omega+\mu)^2 -\left(s_{\perp} \Delta+\sqrt{m^2 +(k^{3})^2}\right)^2}
     +\frac{(\omega-\mu_0)(\omega+\mu+s_{\perp} \Delta)+(k^{3})^2}{(\omega+\mu)^2 -\left(s_{\perp} \Delta-\sqrt{m^2 +(k^{3})^2}\right)^2}
\right]
\nonumber\\
&&-\frac{i}{2(2\pi)^3l^2}\int \frac{d\omega dk^{3}}{\sqrt{m^2 +(k^{3})^2}}
\left[\frac{(\omega-\mu_0)( m )^2+(k^{3})^2(\omega+\mu+s_{\perp} \Delta)}
           {(\omega+\mu)^2 -\left(s_{\perp} \Delta+\sqrt{m^2 +(k^{3})^2}\right)^2}
     -\frac{(\omega-\mu_0)( m )^2+(k^{3})^2(\omega+\mu+s_{\perp} \Delta)}
           {(\omega+\mu)^2 -\left(s_{\perp} \Delta-\sqrt{m^2 +(k^{3})^2}\right)^2}
\right]
\nonumber\\
&-& \frac{i}{(2\pi)^3l^2}\sum\limits_{n=1}^{\infty}
\int d\omega dk^{3}\left[
 \frac{(\omega-\mu_0)(\omega+\mu)+(k^{3})^2+2n|e B |}{(\omega+\mu)^2-\left(E_{k^{3},n}^{+}\right)^2}
+\frac{(\omega-\mu_0)(\omega+\mu)+(k^{3})^2+2n|e B |}{(\omega+\mu)^2-\left(E_{k^{3},n}^{-}\right)^2}
\right]
\nonumber\\
&-& \frac{i}{(2\pi)^3l^2}\sum\limits_{n=1}^{\infty}
\int  \frac{d\omega dk^{3}}{\sqrt{m^2 +(k^{3})^2}} \left[
 \frac{(k^{3})^2 s_{\perp} \Delta}{(\omega+\mu)^2-\left(E_{k^{3},n}^{+}\right)^2}
-\frac{(k^{3})^2 s_{\perp} \Delta}{(\omega+\mu)^2-\left(E_{k^{3},n}^{-}\right)^2}
\right].
\end{eqnarray}
At $T\neq 0$, the result for the free energy density reads
{\small
\begin{eqnarray}
\Omega &=& \frac{T}{2(2\pi l)^2}\int dk^{3}\sum_{m^\prime=-\infty}^{\infty}
\left[\frac{(\omega_{m^\prime}+i\mu_0)(\omega_{m^\prime}-i\mu-i s_{\perp} \Delta)-(k^{3})^2}
           {(\omega_{m^\prime}-i\mu)^2 +\left(E_{k^{3},0}^{+}\right)^2}
     +\frac{(\omega_{m^\prime}+i\mu_0)(\omega_{m^\prime}-i\mu-i s_{\perp} \Delta)-(k^{3})^2}
           {(\omega_{m^\prime}-i\mu)^2 +\left(E_{k^{3},0}^{-}\right)^2}
\right]
\nonumber\\
&-&\frac{T}{2(2\pi l)^2} \int \frac{dk^{3}}{\sqrt{m^2 +(k^{3})^2}}\sum_{m^\prime=-\infty}^{\infty}
\left[\frac{(i\omega_{m^\prime}-\mu_0)m^2+(k^{3})^2(i\omega_{m^\prime}+\mu+s_{\perp} \Delta)}
           {(\omega_{m^\prime}-i\mu)^2 +\left(E_{k^{3},0}^{+}\right)^2}
     -\frac{(i\omega_{m^\prime}-\mu_0)m^2+(k^{3})^2(i\omega_{m^\prime}+\mu+s_{\perp} \Delta)}
           {(\omega_{m^\prime}-i\mu)^2 +\left(E_{k^{3},0}^{-}\right)^2}
\right]
\nonumber\\
&+& \frac{T}{(2\pi l)^2}\sum\limits_{n=1}^{\infty}
\int dk^{3}\sum_{m^\prime=-\infty}^{\infty}\left[
 \frac{(\omega_{m^\prime}+i\mu_0)(\omega_{m^\prime}-i\mu)-(k^{3})^2-2n|e B |}{(\omega_{m^\prime}-i\mu)^2+\left(E_{k^{3},n}^{+}\right)^2}
+\frac{(\omega_{m^\prime}+i\mu_0)(\omega_{m^\prime}-i\mu)-(k^{3})^2-2n|e B |}{(\omega_{m^\prime}-i\mu)^2+\left(E_{k^{3},n}^{-}\right)^2}
\right]
\nonumber\\
&-& \frac{T}{(2\pi l)^2}\sum\limits_{n=1}^{\infty}
\int  \frac{dk^{3}}{\sqrt{ m ^2 +(k^{3})^2}} \sum_{m^\prime=-\infty}^{\infty}
\left[
 \frac{(k^{3})^2 s_{\perp} \Delta}{(\omega_{m^\prime}-i\mu)^2+\left(E_{k^{3},n}^{+}\right)^2}
-\frac{(k^{3})^2 s_{\perp} \Delta}{(\omega_{m^\prime}-i\mu)^2+\left(E_{k^{3},n}^{-}\right)^2}
\right],
\label{OmegaX1}
\end{eqnarray}}
where $E_{k^{3},0}^{\pm} =\sqrt{m ^2 +(k^{3})^2} \pm s_{\perp} \Delta$ and $E_{k^{3},n}^{\pm}
=\sqrt{\left(\sqrt{m ^2 +(k^{3})^2} \pm s_{\perp} \Delta\right)^2 +2n|eB|}$ for $n\geq 1$.
In order to calculate the Matsubara sums in this expression, we used the table sums in
Eqs.~(\ref{tablesum1}) and (\ref{tablesum2}), as well as the following result:
\begin{eqnarray}
X &=& T\sum_{m^\prime=-\infty}^{\infty}
\left(\frac{(\omega_{m^\prime}+i\mu_0)(\omega_{m^\prime}-i\mu)-a^2}{(\omega_{m^\prime}-i\mu)^2+b^2}-1\right)
=T\sum_{m^\prime=-\infty}^{\infty} \frac{(\mu+\mu_0)(i\omega_{m^\prime}+\mu)-a^2-b^2}{(\omega_{m^\prime}-i\mu)^2+b^2}
\nonumber \\
&=&\frac{\mu+\mu_0}{2}\left[n_F(b+\mu)-n_F(b-\mu)\right]-\frac{a^2+b^2}{2b}\left[1-n_F(b+\mu)-n_F(b-\mu)\right].
\end{eqnarray}
Note that the vacuum subtraction in the first line was performed. This is necessary in order to render the
sum finite. In the limit $T\to 0$, the above result reduces to
\begin{eqnarray}
X_{T=0}&=&-\frac{1}{2}\left((\mu+\mu_0) \sign(\mu)
\theta\left(\mu^2-b^2\right)
+\frac{a^2+b^2}{|b|}\theta\left(b^2-\mu^2\right)
\right).
\end{eqnarray}
After performing the Matsubara sums in Eq.~(\ref{OmegaX1}), we obtain
\begin{eqnarray}
\Omega &=& -\frac{1}{2(2\pi l)^2}\int_{0}^{\infty} dk^{3}
\left[
 \frac{(\mu+\mu_0-E_{k^{3},0}^{+})\sinh\left(\mu/T\right)}{\cosh\left(E_{k^{3},0}^{+}/T\right)+\cosh\left(\mu/T\right)}
+\frac{(\mu+\mu_0+E_{k^{3},0}^{-})\sinh\left(\mu/T\right)}{\cosh\left(E_{k^{3},0}^{-}/T\right)+\cosh\left(\mu/T\right)}
\right]
\nonumber\\
&-&\frac{1}{2(2\pi l)^2} \int_{0}^{\infty} \frac{dk^{3}}{E_{k^{3},0}^{+}}
\left[(k^{3})^2+\left(E_{k^{3},0}^{+}\right)^2 +s\Delta(\mu-\mu_0)
+\frac{m^2(\mu-\mu_0)+s\Delta (k^{3})^2}{\sqrt{ m ^2 +(k^{3})^2}}
\right]
\frac{\sinh\left(E_{k^{3},0}^{+}/T\right)}{\cosh\left(E_{k^{3},0}^{+}/T\right)+\cosh\left(\mu/T\right)}
\nonumber\\
&-&\frac{1}{2(2\pi l)^2} \int_{0}^{\infty} \frac{dk^{3}}{E_{k^{3},0}^{-}}
\left[(k^{3})^2+\left(E_{k^{3},0}^{-}\right)^2+s\Delta(\mu-\mu_0)
-\frac{m^2(\mu-\mu_0)+s\Delta (k^{3})^2}{\sqrt{ m ^2 +(k^{3})^2}}
\right]
\frac{\sinh\left(E_{k^{3},0}^{-}/T\right)}{\cosh\left(E_{k^{3},0}^{-}/T\right)+\cosh\left(\mu/T\right)}
\nonumber\\
&-& \frac{1}{(2\pi l)^2}\sum\limits_{n=1}^{\infty}
\int_{0}^{\infty} dk^{3} \left[
 \frac{(\mu+\mu_0)\sinh\left(\mu/T\right)}{\cosh\left(E_{k^{3},n}^{+}/T\right)+\cosh\left(\mu/T\right)}
+\frac{(\mu+\mu_0)\sinh\left(\mu/T\right)}{\cosh\left(E_{k^{3},n}^{-}/T\right)+\cosh\left(\mu/T\right)}
\right]
\nonumber\\
&-& \frac{1}{(2\pi l)^2}\sum\limits_{n=1}^{\infty}
\int_{0}^{\infty}  \frac{dk^{3}}{E_{k^{3},n}^{+}}
\left[
(k^{3})^2+2n|e B |+\left(E_{k^{3},n}^{+}\right)^2
  +\frac{(k^{3})^2 s_{\perp} \Delta}{\sqrt{ m ^2 +(k^{3})^2}}\right]
  \frac{\sinh\left(E_{k^{3},n}^{+}/T\right)}{\cosh\left(E_{k^{3},n}^{+}/T\right)+\cosh\left(\mu/T\right)}
\nonumber\\
&-& \frac{1}{(2\pi l)^2}\sum\limits_{n=1}^{\infty}
\int_{0}^{\infty}  \frac{dk^{3}}{E_{k^{3},n}^{-} }
\left[
(k^{3})^2+2n|e B |+\left(E_{k^{3},n}^{-}\right)^2
-\frac{(k^{3})^2 s_{\perp} \Delta}{\sqrt{ m ^2 +(k^{3})^2}}\right]
\frac{\sinh\left(E_{k^{3},n}^{-}/T\right)}{\cosh\left(E_{k^{3},n}^{-}/T\right)+\cosh\left(\mu/T\right)}.
\label{OmegaX2}
\end{eqnarray}
Note that we used the following identities:
\begin{eqnarray}
n_F(b-\mu)-n_F(b+\mu) = \frac{\sinh(\mu/T)}{\cosh(b/T)+ \cosh(\mu/T)},\\
1-n_F(b-\mu)-n_F(b+\mu) = \frac{\sinh(b/T)}{\cosh(b/T)+ \cosh(\mu/T)}.
\end{eqnarray}

\subsection{Free energy density at $m ,\mu\neq 0$ and $\Delta=\tilde\mu = 0$}

In this case the free energy density (\ref{OmegaX2}) reduces to
\begin{eqnarray}
\Omega &=& -\frac{\mu+\mu_0}{(2\pi l)^2} \int_{0}^{\infty}  dk^{3}\left[
\frac{\sinh\left(\mu/T\right)}{\cosh\left(E_{0}/T\right)+\cosh\left(\mu/T\right)}
+2\sum\limits_{n=1}^{\infty}
\frac{\sinh\left(\mu/T\right)}{\cosh\left(E_{n}/T\right)+\cosh\left(\mu/T\right)}
\right]
\nonumber\\
&-&\frac{1}{(2\pi l)^2} \int_{0}^{\infty}  dk^{3}\left[
\frac{1}{E_{0}}
\frac{\left[(k^{3})^2+E_{0}^2\right]\sinh\left(E_{0}/T\right)}{\cosh\left(E_{0}/T\right)+\cosh\left(\mu/T\right)}
+2\sum\limits_{n=1}^{\infty}
\frac{1}{E_{n}}
\frac{\left[(k^{3})^2+2n|eB|+E_{n}^2\right]\sinh\left(E_{n}/T\right)}{\cosh\left(E_{n}/T\right)+\cosh\left(\mu/T\right)}
\right],
\end{eqnarray}
where $E_{n} =\sqrt{m ^2 +(k^{3})^2+2n|eB|}$ for $n\geq 0$.

In the limit of zero temperature, the above result for the final part of the free energy becomes
\begin{eqnarray}
\Omega &=& -\frac{1}{(2\pi l)^2} \left(\frac{ m ^2}{2}
+\mu_0\sqrt{\mu^2- m ^2}\sign(\mu)\theta(\mu^2- m ^2)
\right)
-\frac{2}{(2\pi l)^2} \sum\limits_{n=1}^{\infty}\left[
\frac{ m ^2}{2}+2n|e B |\ln\frac{\sqrt{2n|e B |}}{\sqrt{ m ^2+2n|e B |}}\right]\nonumber\\
&-& \frac{2}{(2\pi l)^2} \sum\limits_{n=1}^{\infty}\left[
\mu_0\sign(\mu)\sqrt{\mu^2- m ^2-2n|e B |}
+2n|e B | \ln\frac{\sqrt{ m ^2+2n|e B |}}{\sqrt{\mu^2- m ^2-2n|e B |}+|\mu|}
\right]\theta(\mu^2- m ^2-2n|e B |).\nonumber\\
\label{AppAOmega0}
\end{eqnarray}
Note that, in the calculation, we subtracted an infinite constant term:
\begin{equation}
\frac{2}{(2\pi l)^2} \int_{0}^{\infty} dk^{3}\left(k^{3}+2\sum\limits_{n=1}^{\infty} \sqrt{(k^{3})^2+2n|e B |}\right).
\end{equation}
The sum over the Landau levels in the first line of Eq.~(\ref{AppAOmega0}) still contains a logarithmic divergence, i.e.,
\begin{equation}
\Omega_{\rm div}\simeq -\frac{ m ^4}{\left(4\pi\right)^2} \sum_{n=1}^{\infty}\frac{1}{n}
\simeq -\frac{ m ^4}{\left(4\pi\right)^2} \ln (\Lambda l)^2.
\end{equation}
In numerical calculations, we use of the smooth cutoff (\ref{kappa}) to regularize this expression.

\subsection{Free Energy Density at $\Delta,\mu\neq 0$ and $ m =\tilde\mu = 0$}

In this case the free energy density is
\begin{eqnarray}
\Omega &=& -\frac{\mu+\mu_0}{(2\pi l)^2}\int_{0}^{\infty} dk
\left[ \frac{\sinh\left(\mu/T\right)}{\cosh\left(k/T\right)+\cosh\left(\mu/T\right)}
+2\sum\limits_{n=1}^{\infty}
\frac{\sinh\left(\mu/T\right)}{\cosh\left(\sqrt{k^2+2n|eB|}/T\right)+\cosh\left(\mu/T\right)}\right]
\nonumber\\
&-&\frac{1}{2(2\pi l)^2} \int_{0}^{\infty} dk^{3}
\left[2k^{3}+s\Delta +\frac{s\Delta (\mu-\mu_0)}{k^{3}+s\Delta} \right]
\frac{\sinh\left[(k^{3}+s\Delta)/T\right]}{\cosh\left[(k^{3}+s\Delta)/T\right]+\cosh\left(\mu/T\right)}
\nonumber\\
&-&\frac{1}{2(2\pi l)^2} \int_{0}^{\infty} dk^{3}
\left[2k^{3}-s\Delta +\frac{s\Delta (\mu-\mu_0)}{k^{3}-s\Delta} \right]
\frac{\sinh\left[(k^{3}-s\Delta)/T\right]}{\cosh\left[(k^{3}-s\Delta)/T\right]+\cosh\left(\mu/T\right)}
\nonumber\\
&-& \frac{1}{(2\pi l)^2}\sum\limits_{n=1}^{\infty}
\int_{0}^{\infty}  \frac{dk^{3}}{E_{k^{3},n}^{+}}
\left[
(2k^{3} + s_{\perp} \Delta)(k^{3} + s_{\perp} \Delta)+4n|e B | \right]
  \frac{\sinh\left(E_{k^{3},n}^{+}/T\right)}{\cosh\left(E_{k^{3},n}^{+}/T\right)+\cosh\left(\mu/T\right)}
\nonumber\\
&-& \frac{1}{(2\pi l)^2}\sum\limits_{n=1}^{\infty}
\int_{0}^{\infty}  \frac{dk^{3}}{E_{k^{3},n}^{-} }
\left[
(2k^{3} - s_{\perp} \Delta)(k^{3} - s_{\perp} \Delta)+4n|e B |\right]
\frac{\sinh\left(E_{k^{3},n}^{-}/T\right)}{\cosh\left(E_{k^{3},n}^{-}/T\right)+\cosh\left(\mu/T\right)},
\label{OmegaDeltaT}
\end{eqnarray}
where $E_{k^{3},n}^{\pm}=\sqrt{(k^{3} \pm s_{\perp} \Delta)^2+2n|eB|}$.
At $T=0$, after doing the subtraction and the integration, we arrive at
\begin{eqnarray}
\Omega &\simeq & -\frac{\mu_0\mu}{(2\pi l)^2}
   +\frac{s_{\perp} \Delta(\mu_0-\mu)}{(2\pi l)^2} \ln\frac{\Lambda}{|\mu|}\nonumber\\
&-&\frac{2}{(2\pi l)^2} \sum\limits_{n=1}^{\infty} \left[\mu_0\sign(\mu)\sqrt{\mu^2-2n|e B |}
+2n|e B |\ln\frac{\sqrt{2n|e B |}}{|\mu|+\sqrt{\mu^2-2n|e B |}}
\right]\theta\left(\mu^2-2n|e B |\right).
\label{AppAOmegaDelta0}
\end{eqnarray}

\end{widetext}


\begin{thebibliography}{99}
\bibitem{footnote-charge}
Here our convention is such that the electric charge of the electron is $-e$ where $e>0$.
\end{thebibliography}
\end{document}